\documentclass[12pt]{article}
\usepackage{amsmath}
\usepackage{bm}
\usepackage{cite}
\usepackage{mathrsfs}
\usepackage{slashed}
\usepackage{graphicx}
\usepackage{hyperref}
\usepackage{verbatim}
\usepackage{color}
\usepackage{enumerate}
\usepackage{authblk}
\usepackage[hang,flushmargin]{footmisc}
\usepackage{lipsum}
\usepackage[font=footnotesize]{caption}
\usepackage{soul}
\usepackage[utf8]{} 
\usepackage{empheq}
\usepackage{bigints}

\usepackage{amssymb,amsmath,amsfonts}
\usepackage[makeroom]{cancel}
\usepackage[normalem]{ulem}
\usepackage{hyperref}
\usepackage{color}

\newcommand{\diff}{\mathrm{d}}

\numberwithin{equation}{section}

\definecolor{blue-violet}{rgb}{0.54, 0.17, 0.89}
\definecolor{PineGreen}{cmyk}{0.92, 0, 0.59, 0.25}
\definecolor{OliveGreen}{cmyk}{0.64, 0, 0.95, 0.40}
\definecolor{RawSienna}{cmyk}{0, 0.72, 1, 0.45}
\definecolor{Gray}{cmyk}{0, 0, 0, 0.50}
\definecolor{MidnightBlue}{cmyk}{0.98, 0.13, 0, 0.43}
\definecolor{Orange}{cmyk}{0, 0.61, 0.87, 0}
\definecolor{Green}{cmyk}{1, 0, 1, 0}

\renewcommand{\d}{\partial}
\renewcommand{\tilde}{\widetilde}
\renewcommand{\hat}{\widehat}

\def\be{\begin{eqnarray}}
\def\ee{\end{eqnarray}}
\def\beann{\begin{eqnarray*}}
\def\eeann{\end{eqnarray*}}
\def\beq{\begin{equation}}
\def\eeq{\end{equation}}
\def\ba{\begin{array}}
\def\ea{\end{array}}
\def\ben{\begin{enumerate}}
\def\een{\end{enumerate}}
\def\bea{\begin{eqnarray}}
\def\eea{\end{eqnarray}}

\usepackage{atbegshi,picture}
\usepackage{lipsum}
\usepackage[T1]{fontenc}

\def\cH{\mathcal{H}}

\def\dI{\mathscr{I}}

\def\dP{\mathscr{P}}

\usepackage{currvita}

\begin{document}

\title{\vspace{-70pt} \Large{\sc Asymptotic structure of scalar-Maxwell theory at the null boundary} \vspace{10pt}}
\author[a]{\normalsize{Hern\'an A. Gonz\'alez}\footnote{\href{mailto:hernan.gonzalez@uss.cl}{hernan.gonzalez@uss.cl}}}
\author[b]{\normalsize{Oriana Labrin}\footnote{\href{mailto:orianalabrinz@gmail.com}{orianalabrinz@gmail.com}}}
\author[b]{\normalsize{Olivera Miskovic}\footnote{\href{mailto:olivera.miskovic@pucv.cl}{olivera.miskovic@pucv.cl}}}

\affil[a]{\footnotesize\textit{Universidad San Sebasti\'an, Avenida del C\'ondor 720, Santiago, Chile.}}
\affil[b]{\footnotesize\textit{Instituto de F\'isica, Pontificia Universidad Católica de Valparaíso, Avda.~Universidad 330, Curauma, Valparaíso, Chile.}}


\maketitle

\begin{abstract}

We apply the Hamiltonian formalism to investigate the massless sector of scalar field theory coupled with Maxwell electrodynamics through the Pontryagin term. To this end, we generalize the Dirac procedure to include radially independent zero modes of the symplectic matrix associated with the asymptotic symmetries. Specifically, we analyze asymptotic symmetries at the null infinity of this theory, conserved charges, and their algebra. We find that the theory possesses asymptotic shift symmetries of the fields not present in the bulk manifold coming from the zero modes of the symplectic matrix of constraints. Consequently, we conclude that the real scalar field also contains asymptotic symmetries previously found in the literature by a different approach. We show that these symmetries can be seen as the electric-magnetic duality in electromagnetism with the topological Pontryagin term, and obtain a non-trivial central extension between the electric and magnetic conserved charges. Finally, we examine the full interacting theory and find that, due to the interaction, the symmetry generators are more difficult to identify among the constraints, such that we obtain them in the weak-coupling limit. We find that the asymptotic structure of the theory simplifies due to a fast fall-off of the scalar field, leading to decoupled scalar and Maxwell asymptotic sectors, and losing the electric-magnetic duality.

\end{abstract}

\newpage
\tableofcontents

\section{Introduction} 
\label{Introcuction}

The Hamiltonian formalism for constrained gauge systems offers a unique framework for constructing the generators responsible for local and global symmetries. 
However, to correctly perform their identification, it is crucial to provide consistent boundary conditions for the fields at a given boundary and specify the appropriate slicing for the time evolution. 

In \cite{Gonzalez:2023yrz}, we have analyzed the Hamiltonian evolution of electromagnetism and Yang-Mills theory using the retarded (advanced) time slicing. Working in this foliation produces only first  derivatives in the Lagrangian. In the Hamiltonian approach, this is translated to the appearance
of a new constraint that generally does not add new physics. However, once boundary conditions at null infinity are incorporated, it allows us to identify a new type of boundary symmetry. This symmetry produces surface charges that, together with Gauss law, give rise to a centrally extended Kac-Moody algebra. 

In this article, we further investigate the Hamiltonian evolution of gauge theories through null slicing.  Motivated by the analysis of more general deformations and their relation to soft charges, we study the interaction of a gauge field in the presence of a massless scalar through an axionic-like coupling. The action functional of the scalar-Maxwell theory is given by
\be
\label{s.EM}
I[\phi,A] =\frac{1}{2} \int \diff ^4x\,\left(-\sqrt{\mathfrak{g}} \,\mathfrak{g}^{\mu\nu}\partial_{\mu}\phi\partial_{\nu}\phi 
 -\frac{1}{2e^{2}}\sqrt{\mathfrak{g}}\,F_{\mu \nu
}F^{\mu \nu }+\frac{\theta}{4e^2}\,\epsilon ^{\mu \nu \alpha \beta }\,\phi\,F_{\mu \nu
}F_{\alpha \beta } \right)\,,
\ee
where $\mathfrak{g}_{\mu\nu}$ is a four-dimensional background metric. In natural units, the dimensions of the fields $[\phi]$ and $[A]$ are (length)$^{-1}$, and the coupling constants  $[e]=1$ and  $[\theta]=\mathrm{length}$. In the context of celestial holography, similar models have been studied in \cite{Melton:2022fsf,Fan:2022vbz,Stieberger:2023fju,Melton:2023lnz}\footnote{Very recently, a Hamiltonian asymptotic analysis at spatial infinity of electrodynamics non-minimally coupled to scalars has been presented in \cite{Fuentealba:2024mor}.}.

This model provides different scenarios for symmetry enhancement, so we will divide the discussion into three cases. In the first one, we focus on the study of the free massless scalar theory, and we provide a Hamiltonian treatment of the symmetries associated with the scalar soft theorem found in \cite{Campiglia:2017dpg}. 

We then turn to the opposite case, namely, when the scalar field is set to a constant. We obtain Maxwell's electromagnetism endowed with the Pontryagin boundary term.  The Hamiltonian treatment with retarded time allows us to explicitly realize the electromagnetic duality in terms of conjugate pairs of asymptotic surface charges. 

Finally, we consider the most general interaction in \eqref{s.EM}, where gauge fields interact with a massless scalar through a dynamical coupling to the Pontryagin term. It will be seen that electromagnetic duality is lost when consistent boundary conditions are taken into account. However, the infinite-dimensional asymptotic symmetry remains and it is given by the direct sum of the scalar and the photon boundary transformations.

The plan of the paper goes as follows. After the Introduction in Section \ref{Introcuction}, where we motivate and introduce the scalar-Maxwell theory we are interested in, we start our analysis in Section \ref{Formalism} by summarizing the method that will be used to identify symmetry generators, based on the Hamiltonian formalism. We also present the general setup and notation. In Section \ref{Scalar}, we focus only on the sector of the free massless scalar and analyze its asymptotic symmetries. After that, in Section \ref{EM+P}, we consider the electromagnetic sector with the topological Pontryagin term and discuss its asymptotic structure. We particularly examine the electric-magnetic duality in Subsection \ref{Duality}. Finally, we dedicate to the full interacting theory in Section \ref{EM with scalar} and inspect its asymptotic symmetries. We summarize and discuss our results in Section \ref{Discussion}. The technical part related to the derivation of zero modes of the symplectic matrix of the full theory is given in Appendix \ref{Zmodes}.

\section{Local and asymptotic symmetries in canonical formalism}
\label{Formalism}

In the Hamiltonian description of constrained systems, it is not always straightforward to separate first-class constraints (that generate symmetries) from second-class constraints (that eliminate redundant fields), especially in theories with interactions, or non-linear ones as Chern-Simons gauge theory \cite{Banados:1996yj}.

In this section, we provide a summary and a generalization of the usual treatment to separate and identify first-class constraints in a theory with local and/or asymptotic symmetries, following Dirac's
method \cite{dirac2001lectures} (see also \cite{Blagojevic:2002du}). We also introduce basic identities and tools that we will apply frequently in this text.

Let us assume that the system is described by the Hamiltonian
\begin{equation}
\mathcal{H}=\mathcal{H}_{0}+\lambda ^{s}\chi _{s}\,,
\end{equation}
where $\mathcal{H}_{0}$ does not depend explicitly on the constraints. For simplicity, we treat the coordinates $x^i$ that exclude $u$ on the same footing as discrete indices: $(s,x^i) \to s$. 
Here, $\chi _{s}\approx 0$ are all constraints in the theory, and $\lambda^{s}$ are corresponding multipliers.
The symbol `$\approx $' denotes the weak equality, used for the quantities that vanish only on the constraint
surface, and whose Poisson brackets (PBs)  are non vanishing in general. The phase space where all constraints are solved and the canonical variables are reduced only to the independent ones, is called the \emph{reduced phase space}. It is obtained when the full phase space is restricted to the constraint surface.\footnote{On the constraint surface embedded in the phase space, the PBs generate evolution and local transformations. In turn, the Dirac brackets generate the evolution in the reduced phase space, where the constraints are strongly set to zero. This is a standard procedure to eliminate unphysical variables from the phase space. However, applying this standard method without any modification would lose the information about the asymptotic symmetries we are interested in, so we will keep the phase space general until we find all conserved charges.}

In general, the PBs between the constraints are
\begin{equation}
\left\{ \chi _{s},\chi _{s'}\right\} \approx \Omega _{ss'}\,,\label{matrix}
\end{equation}
where $\Omega _{ss'}$ is an antisymmetric symplectic matrix evaluated on the constraint surface. The constraints remain on the constraint surface during their evolution, such that they satisfy
\begin{equation}
\dot{\chi}_{s}\approx \left\{\chi_{s}, \mathcal{H}\right\} \approx 0\quad \Rightarrow \quad \Omega _{ss'}\lambda ^{s'}\approx
Y_{s}\,,  \label{eq.lambda}
\end{equation}
where the non-homogeneous part of the equation is computed from $Y_{s}= \left\{ \mathcal{H}_{0},\chi _{s}\right\} $. The last equation shows that the nature of the multipliers depends only on the symplectic matrix. When the matrix $\Omega $ is invertible (it has maximal rank), there is a unique solution in the multipliers $\lambda =\bar{\lambda}$, and they are fully determined. If the symplectic matrix is equal to zero, then all multipliers are arbitrary, and we denoted them as $\Lambda $.  In the finite-dimensional case, the dimension of the kernel of $\Omega_{ss'} $ specifies the number of independent functions $\Lambda $, while the rank of $\Omega_{ss'} $ determines the number of independent functions $\bar{\lambda}$. However, when the matrix $\Omega_{ss'}(x,x')$ is infinite-dimensional, these numbers are more difficult to count, so the appropriate characterization is given by the dimension of the kernel, the rank of $\Omega$ and the specification of a (sub)manifold where they are defined.

Zero modes $v^{a}$ are obtained from the algebraic equation
\begin{equation}
\Omega _{ss'}V^{s'}=0\,.  \label{z.modes}
\end{equation}
Its nontrivial solutions $V^{s}\neq 0$ have a general form
\begin{equation}
V^{s}=Z^{sa}v_{a}\,,  \label{zv.modes}
\end{equation}%
where $\mathbf{Z}=[Z^{sa}]$ is a known matrix and $v_{a}$ are arbitrary coefficients that correspond to independent zero modes. When the modes are zero, $v_{a}=0$, then all vectors vanish, $V^{s}=0$, and the matrix is invertible.

Since $v_{a}$ are arbitrary,  eqs.~\eqref{z.modes} and \eqref{zv.modes} imply the identity
\begin{equation}
\Omega _{ss'}Z^{s'a}=0\,.
\end{equation}
Then, the general solution of the equation \eqref{eq.lambda} for the
multipliers reads
\begin{equation}
\lambda ^{s}=Z^{sa}\Lambda _{a}+\bar{\lambda}^{s}\,, \label{Zlambda}
\end{equation}
where $\bar{\lambda}^{s}$ is a particular solution of the equation 
\eqref{eq.lambda}, and $\Lambda _{a}$ are arbitrary coefficients, whose number is equal to the number of the zero modes.

With this solution at hand, the constraints in the Hamiltonian become
\begin{equation}
\mathcal{H}=\mathcal{H}_{0}+\lambda _{s}\chi ^{s}=\mathcal{\bar{H}}_{0}+\Xi
^{a}\Lambda _{a}\,,\qquad \Xi ^{a}\equiv \chi _{s}Z^{sa}\,,  \label{FCC}
\end{equation}
where the indefinite part of multipliers ($\Lambda_a$) is associated with the first-class constraints $\Xi ^{a}$ by construction (because $\Lambda_a$ are origin of symmetries), while the determined part $\bar{\lambda}^{s}\chi_{s}$ can be absorbed into the Hamiltonian $\mathcal{\bar{H}}_{0} =\mathcal{H}_{0} +\bar{\lambda}^{s}\chi _{s}$.

Note that the number of first-class constraints $\Xi ^{a}$ is equal to the number of new arbitrary multipliers $\Lambda _{a}$, and equal to the number of the zero modes $v^{a}$, as expected. This means that we can find the first-class constraints in the theory, $\Xi ^{a}$, just by studying the zero modes of $\Omega _{ss'}$ and obtaining the matrix $\mathbf{Z}$.

In practice, we do not need the explicit form of the determined part $\bar{\lambda}^{s}$ of the multipliers because it does not affect the constraints, thus we do not need the explicit form of the functions $Y_{s}$ in \eqref{eq.lambda}.

The smeared generators of local symmetries have the form
\begin{equation}
G[\eta ]=\Xi ^{a}\eta _{a}\,. \label{smeared}
\end{equation}

In special cases when the number of constraints $\chi_s$ is equal to the number of constraints $\Xi_a$, we can use \eqref{FCC} to write the generator as 
\begin{equation}
G[\eta ]=\chi_s\eta^s\,, \qquad  \eta^s \equiv Z^{sa} \eta_a\,, \label{Zpar}
\end{equation}
where $\eta^s$ are new symmetry parameters.
The advantage of the above form is that the information about the zero modes is contained only in the symmetry parameter $\eta^s$, while the constraints remain unchanged. In this way, we can use the original constraints $\chi_s$ as the ones generating the symmetries,  and impose the first-class condition through restrictions on the functional form of the parameter, which is not entirely arbitrary any longer.  This is the case of our interest, as we will see in the next section.

As previously mentioned, all indices have a continuous part, $s\rightarrow (s,x^{i})$, where $x^{i}$ are all coordinates except the retarded/advanced time. This means that, in the calculations above, we neglected all boundary terms and the differentiability of the functional $G[\eta ]$ is not guaranteed. This is resolved via the Regge-Teitelboim method \cite{Regge:1974zd}, by adding a boundary term $Q[\eta ]$ to the smeared generator,
\begin{equation}
G_{Q}[\eta ]=G[\eta ]+Q[\eta ]\,,
\end{equation}
in such a way that the functional derivatives of the improved generator $G_{Q}[\eta ]$ are well-defined. On the constraint surface, the generator
becomes  $G_{Q}[\eta ]\approx Q[\eta ]$, where $Q[\eta]$ is interpreted as a conserved charge using the Noether theorem.
Therefore, on the reduced phase space, we have $G_{Q}[\eta ]=Q[\eta ]$.
In consequence, the constraint algebra becomes, on the constraint surface, the charge algebra, carrying important information about the observables of the theory -- conserved charges and, potentially, centrally extended representations of their algebras.

The classical charge algebra is computed in the canonical representation as
\begin{equation}
\{G_{Q}[\eta _{1}],G_{Q}[\eta _{2}]\}=G_{Q}[[\eta _{1},\eta _{2}]]+C[\eta
_{1},\eta _{2}]\,,
\end{equation}%
where $C[\eta _{1},\eta _{2}]$ is a central extension. On the constraint surface, where $G_{Q}[\eta ]=Q[\eta ]$ and the PBs are replaced by the Dirac brackets $\{\; ,\; \}^*$, it becomes
\begin{equation}
\{Q[\eta _{1}],Q[\eta _{2}]\}^{\ast }=Q[[\eta _{1},\eta _{2}]]+C[\eta
_{1},\eta _{2}]\,.
\end{equation}
However, to compute the above algebra, we will use the symplectic differential geometry (explained and applied in Section \ref{EM+P}), which can arrive at the above result valid on the reduced phase more directly, without the need to introduce Dirac brackets explicitly.

We will use the above method to analyze the symmetries and charge algebra in the scalar-Maxwell theory \eqref{s.EM} on the background of the null-foliated spacetime.\medskip

In the next section, we detail our conventions and the coordinate system used throughout this article.

\subsection{Setup and notation}

To study the effects on null boundaries of field theories more systematically, we work in the four-dimensional Minkowski space background written in Bondi's coordinates $x^\mu=(u,r,y^A)$ as
\begin{eqnarray}
\label{flat}
\diff s^{2} &=& \mathfrak{g}_{\mu\nu}(x)\,\diff x^\mu \diff x^\nu
=-\diff u^{2}-2\epsilon\, \diff u \diff r+r^{2}\diff \Omega ^{2}\,.
\end{eqnarray}
Here, $r\geq 0$ is the radial coordinate and $y^A$ are  angles of the unit 2-sphere with the line element $\diff \Omega ^{2}=\gamma _{AB}(y )\,\diff y ^{A}\diff y ^{B}$. 
We also introduced the retarded time, $\epsilon=1$, and advanced time, $\epsilon=-1$, related to the usual time $t$ as $u=t-\epsilon r\in \mathbb{R}$. In that way, we have an explicit realization of the null-foliated flat spacetime, with the waves ($\diff s=0$) propagating along the lines $u=\mathrm{const.}$~and $u+2\epsilon r=\mathrm{const}$.
The spacetime manifold has the topology $M \cong \mathbb{R}\times \Sigma $, valid at least locally, where $\Sigma $ is the 
null surface parametrized by
\footnote{The null coordinates $\ell _{\mu }\mathrm{d}x^{\mu }=\mathrm{d}u$ and $n_{\mu }\mathrm{d}x^{\mu }=\mathrm{d}u+2\epsilon
\,\mathrm{d}r$  define the null vectors $\ell _{\mu }=\partial _{\mu
}u=(1,0,0,0)$ and $n_{\mu }=(1,2\epsilon ,0,0)$. Thus, the $u=\mathrm{const}.
$ surface is lightlike, i.e., with the normal vector $\ell _{\mu }$ satisfying $g^{\mu \nu }\ell _{\mu }\ell _{\nu }=0$.}
$x^i=(r,y^A)$. Chosen coordinates correspond to the future cone when $\epsilon=1$, and to the past cone when $\epsilon=-1$.

For completeness, the flat metric and the Jacobian in the Bondi coordinates have the form 
\begin{equation}
\mathfrak{g}_{\mu \nu }=\left(
\begin{array}{ccc}
-1 & -\epsilon & 0 \\
-\epsilon & 0 & 0 \\
0 & 0 & r^{2}\gamma _{AB}
\end{array}
\right) ,\quad \mathfrak{g}^{\mu \nu }=\left(
\begin{array}{ccc}
0 & -\epsilon & 0 \\
-\epsilon & 1 & 0 \\
0 & 0 & \frac{1}{r^{2}}\,\gamma ^{AB}
\end{array}
\right) ,\quad \sqrt{\mathfrak{g}}=r^{2}\sqrt{\gamma }\,,  \label{metric}
\end{equation} 
where  $\mathfrak{g}=|\det[\mathfrak{g}_{\mu \nu}]|$ and $\gamma=|\det[\gamma_{AB}]|$ are the absolute values of the determinants of $\mathfrak{g}_{\mu \nu}(x)$ and $\gamma_{AB}(y)$. The metric $\gamma _{AB}$ and its inverse $\gamma ^{AB}$ raise and lower the indices on the 2-sphere. In the case of four-dimensional vectors,  we use the notation
\begin{equation}
   B^\mu=\mathfrak{g}^{\mu \nu}B_\nu\,,\quad \tilde{B}^A=\gamma^{AB}B_B \quad \Rightarrow \quad B^A=\frac{1}{r^2}\, \tilde B^A \,. \label{tilde}
\end{equation}
 For the sake of simplicity, we will not write the tilde when raising/lowering the indices of two-dimensional vectors that do not have a four-dimensional form.

\section{Scalar field theory}
\label{Scalar}

The particular features of evolution in retarded time can already be understood in the simplest case of a massless scalar field in the Bondi background \eqref{flat}. We will see that this system possesses an infinite set of Abelian charges. These generators coincide with the ones previously obtained in \cite{Campiglia:2017dpg,Campiglia:2018see}, motivated by the existence of a scalar soft theorem. See also \cite{Simic:2023exz} for a treatment of the massless scalar symmetries in double null coordinates.

We consider massless matter fields in the Minkowski space background \eqref{metric}. We focus first on the action \eqref{s.EM} without the electromagnetic field because we expect non-trivial asymptotic symmetries even in this simple setting. Thus, taking $A_\mu=0$, the action is given by
\begin{equation}
    I_{\mathrm{S}}[\phi]= - \frac{1}{2}\int \diff ^4x\,\sqrt{\mathfrak{g}} \,\mathfrak{g}^{\mu\nu}\partial_{\mu}\phi\partial_{\nu}\phi  \,. \label{scalar}
\end{equation}
We will analyze how the system evolves along the null direction $u$, thus we denote the scalar field velocity by $\partial_{u}\phi=\dot\phi$. As expected in the null foliation, the scalar field action becomes a functional linear in velocities \cite{Dirac:1949cp, Alexandrov:2014rta}, leading to the velocity-independent canonical momenta 
$    \pi = \epsilon r^2\sqrt{\gamma}\,\partial_{r}\phi$,
and the primary constraint
\begin{equation}
    \chi = \epsilon\pi - r^2\sqrt{\gamma}\,\partial_{r}\phi\approx 0\,. \label{chi.S}
\end{equation}
 The fundamental canonical PBs have the form
\begin{equation}
\left\{ \phi ,\pi'\right\} =\delta ^{(3)}(x-x')\,,
\end{equation}
with the shorthand notation $\phi \equiv \phi(u,r,y)$ and $\pi' \equiv \pi(u,r',y')$, and the Dirac delta function $\delta ^{(3)}$ is defined on the  null surface with constant $u$  in spherical coordinates.
The constraint \eqref{chi.S} is not present when the evolution is studied in the usual timelike foliation of spacetime. Its appearance is a special feature of the time being null. Besides, the PBs between the primary constraints,
\begin{equation}
    \{ \chi,\chi ' \}\equiv \Omega(x,x') = -2\epsilon r\sqrt{\gamma}\,\left(1+r\partial_r\right) \delta^{(3)}\,,
\end{equation}
define the symplectic matrix $\Omega$ that does not have a unique inverse, because of the zero mode $v(u,y)$ obtained from the equation \eqref{z.modes} as
\begin{equation}
\int \diff^{3}x'\,\Omega(x,x')V(x')= -2\epsilon \sqrt{\gamma } \partial _{r}\left( rV(x)\right)=0 \quad\Rightarrow\quad V(x)= \frac{v(u,y)}{r}\,.
\end{equation}
Therefore, the kernel of the matrix $\Omega$ has dimension one and its rank is zero in the subspace $r=\mathrm{const}.$, while its rank is one in the full spacetime.
We identify the matrix $Z$ in \eqref{zv.modes} as $1/r$. The fact that the zero mode appears at the order $1/r$ in the radial expansion suggests that the first-class constraint and the corresponding symmetry will also become relevant at the order $1/r$ of the local parameter, as suggested by \eqref{Zpar}.

\paragraph{Hamiltonian and  Hamilton's equations.} 

Knowing the momenta and the primary constraint, we can write the total Hamiltonian density in the form
\begin{equation}
\mathcal{H}_T= \frac{1}{2}\,r^2\sqrt{\gamma}\,(\partial_r \phi)^2 + \frac{1}{2}\sqrt{\gamma}\,\gamma^{AB}\partial_{A}\phi\partial_{B}\phi + \lambda\chi\,,
\end{equation}
where the function $\lambda (x)$ is a multiplier. The Hamiltonian $H_T=\int \diff^3 x \, \mathcal{H}_T $ governs the evolution of phase space functions $\mathcal{F}(\phi,\pi,\lambda)$ as $\dot{\mathcal{F}}\approx\int \diff^3 x' \,\{\mathcal{F}, \mathcal{H}'_T \}$.

The multiplier is partially determined by the requirement
that the constraint remains on the constraint surface during its evolution,
\begin{equation}
\dot{\chi}=\int \mathrm{d}^{3}x'\,\left\{ \chi ,\mathcal{H}'_T\right\} =\epsilon\partial _{j}\left( \sqrt{\mathfrak{g}}\,\mathfrak{g}^{ij}\partial_{i}\phi \right) -2\epsilon r\sqrt{\gamma}\,(1+r\partial _{r})\lambda  =0\,, 
\end{equation}
which leads to a differential equation for the multiplier  
\begin{eqnarray}
  \partial_r(r\lambda)&=&\frac{1}{2r}\,\partial _{r}\left( r^{2}\partial _{r}\phi \right) + 
\frac{1}{2r} \,\tilde\Box\phi\,,
\end{eqnarray}
where $\tilde\Box\phi = \frac{1}{\sqrt{\gamma}}\, \partial_A\left( \sqrt{\gamma}\,\gamma^{AB}\partial_B\phi\right)$ is the D'Alembert operator on the 2-sphere.
The general solution for the multiplier has the form 
\begin{equation}
\lambda =\bar{\lambda}+\frac{\Lambda (u,y)}{r}\,,
\end{equation}
with the determined part
\begin{equation}
\bar{\lambda}=-\int\limits_{r}^{\infty }\frac{\mathrm{d}r}{r^{2}}\,\left[
\tilde{\square}\phi +\partial _{r}(r^{2}\partial _{r}\phi )\right] .
\end{equation}
This form of the multiplier $\lambda$ is consistent with \eqref{Zlambda} because it has already been found that $Z=1/r$.
The indefinite part $\Lambda (u,y)$ is associated with the zero modes $v(u,y)$ of the symplectic matrix. Indeed, they appear in the same order. The existence of $v$ introduces arbitrariness in the evolution of the system, which indicates that there is a symmetry.

The Hamilton's equations for the scalar field and its momentum are 
\begin{equation}
\dot{\phi}= \epsilon\lambda \,, \qquad
\dot{\pi} = \sqrt{\gamma}\,\partial_r(r^{2}\partial_r\phi) + \sqrt{\gamma}\,\tilde\Box\phi-\sqrt{\gamma}\,\partial_r(\lambda r^2)\,.\label{eqs.S}
\end{equation}
We note that the time evolution of the scalar field is encoded in the multiplier \cite{Alexandrov:2014rta}.

The fall-off of the fields in agreement with Hamilton's equations are 
\begin{equation}
\phi =\mathcal{O}(r^{-1})\,,\qquad \pi =\mathcal{O}(r^{0})\,, \qquad \lambda =\mathcal{O}(r^{-1})\,,  \label{fall.S}
\end{equation}
and they keep this form even when an interaction is introduced \cite{Campiglia:2018see}.

\paragraph{Asymptotic symmetry and charges.} 

Since there exists a non-trivial zero mode of the symplectic matrix, the corresponding first-class constraint in $\chi$  generates an asymptotic symmetry. To find this constraint, we introduce a smeared generator defined on the null manifold according to \eqref{Zpar} as 
\begin{equation}
K[\xi]=\int \mathrm{d}^{3}x\,\xi \chi\,, \qquad  \xi=\frac{1}{r}\,\xi_{(1)}(u,y)\,. \label{smear.S}
\end{equation}
Now the local parameter carries the information about the zero mode of the symplectic matrix, according to \eqref{Zpar}. This form of the local parameter $\xi$ is essential, as it ensures that only first-class constraints contribute to the smeared generator.  Indeed, such generators close an Abelian algebra,
\begin{eqnarray}
\left\{ K[\zeta ],K'[\xi' ]\right\}  &=&\int \mathrm{d}^{3}x\int
\mathrm{d}^{3}x'\,\zeta \xi'\,\Omega(x,x')  \notag \\
&=&-2\epsilon \int \mathrm{d}^{3}x\sqrt{\gamma }\,\zeta _{(1)}\left(\rule[2pt]{0pt}{9pt}
1+r\partial _{r}\right) \frac{\xi _{(1)}}{r}   =0\,.
\label{Genesc}
\end{eqnarray}
It implies that $K$ is a symmetry generator. Since the parameter $\xi_{(1)}$ is $r$-independent, $K$ is a generator of a symmetry that is evident in the asymptotic sector via the spherical harmonics coefficients $\xi_{(1)\ell m}$ on the celestial sphere. This will give rise to transformations of the fields at the boundary.

In the context of gauge theories, gauge transformations can have boundary conditions on the parameters that do not decay at infinity, thus converting the gauge transformations into improper ones \cite{Benguria:1976in}. For the massless scalar field, however, we do not have a local symmetry in the bulk, and the obtained symmetry arises only on the boundary \cite{Campiglia:2017dpg}. Note that we have imposed conditions on the parameter $\xi$ such that it is finite at infinity.

To be differentiable, the smeared generator has to be supplemented by a boundary term $Q_K$, such that the full generator is 
\begin{equation}
G_K[\xi ]=\int \mathrm{d}^{3}x\,\xi \chi+Q_K[\xi ]\approx Q_K[\xi ]\,.
\end{equation}
The above equation shows that the generator becomes $G_K[\xi ]=Q_K[\xi ]$ on the reduced phase space.
Since, on the constraint surface, $Q_K[\xi ]$ generates symmetry transformations, $Q_K$ can be identified with the conserved charge. It is obtained from the conditions that the functional derivatives $\delta G_K[\xi ]/\delta \phi$ and $\delta G_K[\xi ]/\delta \pi$ are well-defined, leading to
\begin{equation}
Q_K[\xi ]= r^{2}\oint\mathrm{d}^{2}y\,\sqrt{\gamma }\,\xi \phi \,,
\label{charge.s}
\end{equation} 
when $\xi$ is field-independent ($\delta \xi=0$). The integral is defined on the celestial sphere $\mathbb{S}^2$ located at the boundary $r \to \infty$. As regards the time dependence, it is specified as follows. 

For a radiating massless system, the charge is defined as the total flux  $\int \mathrm{d}u\,\partial _{u}Q_{K}$ \cite{Sachs:1962wk} over the boundary that, in the future cone ($\epsilon =1$), corresponds to  $-Q_{K}|_{\dI_{-}^{+}}$ in absence of massive particles in the spectrum ($Q_{K}|_{\dI_{+}^{+}}=0$). This means that the charge \eqref{charge.s} cannot be evaluated at arbitrary time $u=u_{0}=\mathrm{const}$, but at $u \to -\infty$. This is secured if 
one completes the integral \eqref{smear.S} by
adding a patch of null infinity from $u_{0}$ to $u\rightarrow -\infty $. The final charge $Q^+$ defined only over the boundary of the future null infinity is 
\begin{equation}
Q^+[\xi ]=-r^{2}\oint_{\dI_{-}^{+}}\mathrm{d}^{2}y\,\sqrt{\gamma } \,\xi \phi \,.
\label{charge+.s}
\end{equation}
Doing so, we also ensure that the integral is always performed over a Cauchy surface (see \cite{Donnelly:2016auv,Geiller:2017xad,Geiller:2017whh,Hosseinzadeh:2018dkh,Speranza:2017gxd} for a deeper discussion).  In practice, it
means that all boundary quantities in the charge are $u$-independent
(evaluated at $u\rightarrow -\infty $). Because $\xi_{(1)}$ does not depend on $r$, it implies that it is $u$-independent in the whole spacetime, namely, $\xi=\frac{\xi_{(1)}(y)}{r}$. The charge $Q^-$ similar to \eqref{charge+.s} can be defined in the past cone ($\epsilon=-1$), with the global sign positive and the integral evaluated at $\dI_{+}^{-}$, that is, in the limit $u \to +\infty$. In sum, $Q^\pm[\xi]=-\epsilon \, Q_K[\xi(y)]\,|_{u\to -\epsilon \infty}$.

To simplify the notation for all $\epsilon=\pm 1$, we will work with the usual Hamiltonian charge \eqref{charge.s}, assuming the $u$-independence of the asymptotic symmetry parameter and with the boundary integral evaluated at $\dI_{+}^{-}/\dI_{-}^{+}$. Differences in signs can always be absorbed in the local parameter.

The transformation law of any function $\mathcal{F}(\phi,\pi)$ on the reduced  phase space is generated through PBs as $\delta _{\xi }\mathcal{F}=\left\{ \mathcal{F},G_K[\xi ]\right\}$, such that the transformation law of the canonical fields reads
\begin{equation}
\delta _{\xi }\phi =\epsilon \,\xi \,,\qquad \delta _{\xi }\pi = -\partial_r(r^2 \sqrt{
\gamma }\,\xi\,).
\label{transf1.S}
\end{equation}
We recognize the shift symmetry of the scalar field, the way Goldstone particles change in the vacuum. 
The fall-off of the parameter $\xi=\mathcal{O}(r^{-1})$, as defined by \eqref{smear.S}, is consistent with the above laws. For instance, the invariance of Hamilton's equation for the momentum  \eqref{eqs.S} is satisfied only in the asymptotic limit, leading to $\dot{\xi}\rightarrow 0$ when $r\rightarrow \infty$, which is consistent with the $u$-independence of $\xi $. Furthermore, invariance of the first Hamilton's equation \eqref{eqs.S} under these transformations and the fact that the parameter is $u$-independent, imply invariance of the Hamilton's multiplier under symmetry transformations,
\begin{equation}
   \delta_{\xi}\lambda= \dot{\xi} =0\,, \label{transf2.S}
\end{equation}
while the second equation is satisfied asymptotically in the leading order without any restriction on the parameter. If $\mathcal{F}_{(n)}$ 
denotes an order $r^{-n}$ in the large $r$ expansion of $\mathcal{F}$, the only components of the fields that change under transformations  \eqref{transf1.S} and \eqref{transf2.S} at the asymptotic boundary $r \to \infty$, are their leading orders, 
\begin{equation}
\delta _{\xi}\phi_{(1)} =\epsilon\, \xi_{(1)}\,,\qquad \delta _{\xi}\pi_{(0)} = -\sqrt{\gamma}\,\xi_{(1)}\,, \qquad \delta_{\xi}\Lambda =0\,,
\end{equation}
and the conserved charge $Q_K[\xi]$ is indeed finite,
\beq
Q_K[\xi_{(1)}]=\oint \diff ^2y \, \sqrt{\gamma} \, \xi_{(1)}\phi_{(1)}  \,. \label{QK}
\eeq

As discussed previously, the conserved quantity for a radiating system in the future cone coincides with the Hamiltonian charge  \eqref{QK} evaluated at the past of the future light-like infinity, $Q^+=-Q_K|_{\dI^{+}_{-}}$. This charge also matches the one discussed in \cite{Campiglia:2017dpg} concerning the soft theorem for the scalar field. Similarly, the conserved quantity for a radiating system in the past cone is the Hamiltonian charge evaluated at the future of the past light-like infinity, $Q^-=Q_K|_{\dI^{-}_{+}}$.
From the total flux, then, we can write the usual conservation law of the massless system,
$Q^+=Q^-$, provided suitable matching conditions. 
The antipodal matching conditions $\phi _{(1)}(u\to -\infty ,y)|_{\epsilon=1}=\phi _{(1)}(u \to \infty ,-y)|_{\epsilon=-1}$ and $\xi (y)|_{\epsilon =1}=-\xi (-y)|_{\epsilon =-1}$ (the last one due to our definition of the parameter given by
eq.~\eqref{transf1.S}) satisfy continuously the asymptotic massless Klein-Gordon equation in the matching points at the null infinity, which is in agreement with \cite{Henneaux:2018mgn,Campiglia:2018see,Campiglia:2017dpg}.

Finally, it can be shown that the charge algebra in the reduced phase space is Abelian,
\begin{equation}
    \{ Q_K[\xi_1],  Q_K[\xi_2] \}^{\ast } =0\,,
\end{equation}
without a central extension. 
The next section will show more details about the computation of these algebras.

\section{Electromagnetism with the Pontryagin term}
\label{EM+P}

Having analyzed the pure scalar field symmetries of \eqref{s.EM} in the previous section, now we can focus on the other extreme and assume the constant scalar field, $\phi=\phi_0=\mathrm{const}$. This constant can be absorbed into the coupling parameter, $\theta\phi_0 \to \theta$. We obtain the Maxwell's action with the Pontryagin term  in the  background $\mathfrak{g}_{\mu \nu}$, 
\begin{equation}
I[A]=\frac{1}{4e^{2}}\int \diff^{4}x\left( -\sqrt{\mathfrak{g}}\,F_{\mu \nu
}F^{\mu \nu }+\frac{\theta}{2}\,\epsilon ^{\mu \nu \alpha \beta }\,F_{\mu \nu
}F_{\alpha \beta }\right) \,, \label{Maxwell}
\end{equation}
where $A_{\mu}(x)$ is the gauge field with the associated field strength $ F_{\mu \nu}=\d_\mu A_{\nu}-\d_\nu A_\mu$, and  $\theta$ and $e$ are dimensionless coupling  constants.\footnote{Note that the replacement $\theta\phi_0 \to \theta$ changes the dimension of the coupling $\theta$.} Asymptotic symmetries of the pure Maxwell action were analyzed in the Hamiltonian formalism at null infinity in \cite{Gonzalez:2023yrz} and \cite{Bunster:2018yjr}. The Pontryagin term, whose coupling is $\theta$, is a Lorentz and gauge invariant topological invariant, which does not change Maxwell's equations. Namely, the Pontryagin density is, locally, a total divergence,
\beq
\frac{1}{4} \,\epsilon^{\mu\nu\alpha\beta} \,F_{\mu \nu} F_{\alpha\beta} = \partial_{\mu} \left(\epsilon^{\mu\nu\alpha\beta}A_{\nu}\partial_{\alpha} A_{\beta} \right)    \,,
\eeq
but globally, when properly normalized, it can give rise to a nontrivial Pontryagin index, depending on the topology of spacetime.

The canonical momenta are defined by
\begin{equation}
\pi ^{\mu} =-\frac{%
\sqrt{\mathfrak{g}}}{e^{2}}\,F^{u\mu }+\frac{\theta}{2e^{2}}\,\epsilon ^{u\mu
\alpha \beta }F_{\alpha \beta }\,.
\end{equation}
Thus, adding the topological term in the action is equivalent to having a $\theta$-dependent shift in the canonical momenta. It is electromagnetic analogous to Ashtekar's variables \cite{Ashtekar:1986yd}.

The canonical PBs have the usual form,
\beq
\{A_{\mu}(x),\pi^\nu(x')\}=\delta^{\mu}_{\nu} \delta^{(3)}(x-x')\,.
\eeq
To identify the constraints, we use the explicit form of the background metric \eqref{flat}, and the momenta components become
\begin{equation}
\pi ^{u} =0\,,\qquad \pi ^{r}=\frac{r^{2}}{e^2}\,\sqrt{\gamma }
F_{ur}+\frac{\theta}{2e^{2}}\,\epsilon ^{AB}F_{AB}\,,  \qquad
\pi ^{A} = \epsilon \sqrt{\gamma }\,\sigma
^{AB}F_{rB}\,. \label{pi.EM}
\end{equation} 
For the reduction of the Levi-Civita symbol to two dimensions, we use the notation $\epsilon^{urAB}\equiv \epsilon^{AB}$.   We also introduce the invariant tensor 
\begin{equation}
\sigma ^{AB}=\frac{1}{e^{2}}\,\left( \gamma ^{AB}-\frac{\epsilon \theta }{\sqrt{\gamma }}\,\epsilon ^{AB}\right) \,,
\end{equation}
whose transpose tensor corresponds to the replacement $\theta \to -\theta$,
and they satisfy the property 
\begin{equation}
(\sigma \sigma ^{T})_{\ B}^{A}=\sigma^{AC} \sigma_{BC} =\frac{1+\theta^2}{e^4}\, \delta _{B}^{A}\,. \label{sigma^2}
\end{equation}
In the computations, we use the identity
\begin{equation}
\epsilon ^{AB}\epsilon _{AC}=\delta _{C}^{B}\,,
\end{equation}
fulfilled when the 2-sphere is parametrized by the spherical angles $y^{A}=(\vartheta ,\varphi )$ .\footnote{In the complex coordinates $(z,\bar{z})$ of the Riemann sphere, the signature changes and the identity becomes $\epsilon ^{AB}\epsilon
_{AC}=-\delta _{C}^{B}$.}

The primary constraints are
\begin{equation}
\pi^u\approx 0\,,\qquad \chi^{A}=\epsilon \pi
^{A} -\sqrt{\gamma}\,\sigma^{AB} F_{rB}\approx 0\,. \label{prim.EM}
\end{equation}
Similarly as in the case without the Pontryagin term \cite{Gonzalez:2023yrz}, the first primary constraint is related to the $\mathrm{U}(1)$ gauge symmetry, while the second one is characteristic for the null foliation of spacetime and it will give rise to an asymptotic symmetry. 

The property \eqref{sigma^2} can be used to invert the above relation and solve the field strength in terms of the momenta as
\begin{equation}
F_{rA}=\frac{e^4}{(1+\theta^2) \sqrt{\gamma }}\, \sigma _{BA}\left( \epsilon \pi^{B}-\chi^{B}\right) \,. \label{inverse}
\end{equation}

Now we can compute the total Hamiltonian, using the identity\footnote{Covariant versions of the rank-2 Levi-Civita symbols are $\frac{1}{\sqrt{\gamma }}\,\epsilon
^{AB}$ and $\sqrt{\gamma }\,\epsilon _{AB}$. Then, the contraction results in
$\frac{1}{\gamma }\,\epsilon ^{AB}\epsilon ^{CD}F_{AB}F_{CD}=\left(
\gamma ^{AC}\gamma ^{BD}-\gamma ^{AD}\gamma ^{BC}\right)
F_{AB}F_{CD}=2\tilde{F}^{AB}F_{AB}$.}
\begin{equation}
\frac{1}{2\gamma }\,\epsilon ^{AB}\epsilon ^{CD}F_{AB}F_{CD}=\tilde{F}^{AB}F_{AB}\,.
\label{id}
\end{equation}
We applied the notation \eqref{tilde}, such that $\tilde{F}^{AB}=\gamma ^{AC}\gamma
^{BD}F_{CD}$. This quantity differs from the four-dimensional tensor $F^{\mu\nu }$ which, projected to the 2-sphere, has the form $F^{AB}=r^{-4}\,\tilde{F}^{AB}$.

\paragraph{Hamiltonian and equations of motion.}

By means of \eqref{sigma^2}, the canonical Hamiltonian density, obtained as the Legendre transformation of the Lagrangian density, has the form
\begin{equation}
\mathcal{H}_{C}=\frac{e^{2}}{2r^{2}\sqrt{\gamma }}\,\pi^{r}\left( \pi ^{r}-\frac{\theta}{e^2}\,\epsilon ^{AB}F_{AB}\right) +\frac{\sqrt{\gamma }\gamma ^{AB}}{2e^{2}}\,F_{rA}F_{rB}+\frac{\sqrt{\gamma}}{4e^2r^{2}}\,(1+\theta^2) \,F_{AB}\tilde{F}^{AB}-A_{u}\d_{i}\pi^{i}\,,
\end{equation}
where we neglected the total divergence $\d_i(A_u\pi^i)$ that will become a boundary term in the Hamiltonian. 

The total Hamiltonian density includes all primary constraints, so we introduce the multipliers $\lambda _{u}$ and $\lambda _{A}$ associated to the constraints
\begin{equation}
\cH_{T}=\cH_{C}+\lambda_A \chi^A+ \lambda_u \pi^u \,. 
\end{equation}
The above Hamiltonian is equivalent to the one written in \cite{Gonzalez:2023yrz}, up to the term $\gamma ^{AB}F_{rA}F_{rB}$ which was previously written in terms of the momenta as $\pi^A\tilde\pi_A $, with the price that the old multiplier had to be shifted by a field-dependent vector, $\lambda _{A} \to \lambda _{A}-\frac{e^2 }{1+\theta^2 }\,\sigma _{A}^{\ B}F_{rB}$. 

Hamilton's equations for momenta read
\begin{eqnarray}
\dot{\pi}^{u} &\approx &  0\,, \qquad
\dot{\pi}^{r} =\sqrt{\gamma }\,\nabla ^{A}\left( -\frac{1}{e^{2}}
\,F_{rA}+\lambda ^{B}\sigma _{BA}\right) \,,  \notag \\
\dot{\pi}^{A} &=&\frac{1}{r^{2}}\,\partial _{B}\left( \theta \,\epsilon
^{AB}\pi ^{r}-\frac{1+\theta^2}{e^2 }\,\sqrt{\gamma }\,\tilde{F}^{AB}\right) +\sqrt{\gamma }\,\partial _{r}\left( \frac{1}{e^{2}}\,\gamma
^{AB}F_{rB}-\sigma ^{AB}\lambda _{B}\right) \,, \label{eomP.EM}
\end{eqnarray}
and, for the gauge fields, they read
\begin{eqnarray}
\dot{A}_{u} &=&\lambda _{u}\,,  \qquad
\dot{A}_{r} =\frac{e^{2}}{r^{2}\sqrt{\gamma }}\,\left( \pi ^{r}-\frac{\theta}{2e^2}
\,\epsilon ^{AB}F_{AB}\right) +\partial _{r}A_{u}\,,  \qquad
\dot{A}_{A} =\partial _{A}A_{u}+\epsilon \lambda _{A}\,. \label{eomA.EM}
\end{eqnarray}
They depend explicitly on $\theta$, even though the Pontryagin term is a topological invariant that does not change the bulk dynamics of the theory. The reason is a $\theta$-dependent shift in the definition of momenta. It can be shown that the dynamics remain unchanged because the 
second-order Euler-Lagrange equations obtained from the above Hamilton's equations cancel out all the $\theta$ terms. Thus, the bulk physics is unchanged, but the Pontryagin term can produce nontrivial boundary effects.

\paragraph{Evolution of constraints.} 

Requiring that the retarded time evolution keeps the constraints \eqref{prim.EM} on the constraint surface, we find that $\pi^u$ leads to a secondary constraint, the Gauss law 
\begin{equation}
\dot{\pi}^{u}\approx 0\quad \Rightarrow \quad \chi = \partial_{i}\pi
^{i}\approx 0\,,
\end{equation}
which also appears explicitly in the Hamiltonian as  $\mathcal{H}_T=\mathcal{H}_{0}-A_{u}\chi$. The component $A_u$ has a weakly zero associated conjugate momentum. Therefore, it enters linearly, behaving as a  Hamiltonian multiplier associated to $\chi$. Thus, the total Hamiltonian can conveniently be written as 
\begin{equation}
\cH_{T}=\mathcal{H}_{0}-A_{u}\chi+\lambda_A \chi^A+ \lambda_u \pi^u \,.
\end{equation}
Non-vanishing PBs between the constraints are
\begin{equation}
\left\{ \chi ^{A},\chi^{\prime B}\right\}  =\Omega^{AB}(x,x')\,,  \qquad
\left\{ \chi^{A},\chi'\right\}  =0\,, \qquad
\left\{ \chi,\chi'\right\}  = 0\,,  
\end{equation}
where 
\begin{equation}
\Omega^{AB}(x,x')=-\frac{2\epsilon }{e^{2}}\,
\sqrt{\gamma }\,\gamma ^{AB}\partial _{r}\delta^{(3)}
\end{equation}
is a $\theta $-independent symplectic form. 
In all the computations, we use the formula that applies to any tensor $\mathcal{F}$,
\begin{equation}
\int \mathrm{d}^{3}x'\,\mathcal{F}'\,\left\{ \pi ^{\mu },
F'_{\alpha \beta }\right\} =\d_{\alpha }\mathcal{F}\,
\delta _{\beta }^{\mu }-\d_{\beta }\mathcal{F}\,\delta _{\alpha }^{\mu
} \,. \label{id.EM}
\end{equation}

It is straightforward to show that the constraint $\chi$ remains vanishing during its evolution,   while the evolution of $\chi^{A}$ partially determines the associated multiplier from the linear differential equation 
\begin{equation}
\dot{\chi}^{A}\approx \int \diff^{3}x'\left\{ \chi^{A},\mathcal{H}'_{0}\right\} +\int \diff^{3}x'\,\Omega^{AB}\lambda'_{B}\approx 0\quad \Rightarrow\quad \d_{r}\lambda _{A}\approx -\frac{\epsilon e^2}{2\sqrt{\gamma}}\,Y_{A}\,.
\end{equation}
The inhomogeneous part of this equation, $Y_{A}$, can be computed from $\int \diff^{3}x'\left\{ \mathcal{H}'_{0},\chi_A\right\}$. We conclude that a solution of this equation has an `integration constant',  an $r$-independent function $\Lambda_A (u,y)$. It is a direct consequence of the fact that, again, the symplectic form $\Omega^{AB} $ has an $r$-independent zero mode, given by \eqref{z.modes},
\begin{equation}
\int \mathrm{d}^{3}x'\,\Omega ^{AB}(x,x')V_{B}(x')=0\quad \Rightarrow \quad V_{A}=v_{A}(u,y)\,.
\end{equation}
This zero mode is the same as in Maxwell's electromagnetic theory \cite{Gonzalez:2023yrz}, because $\Omega^{AB} $ is $\theta $-independent. Its geometric interpretation is the same as in the scalar field theory; namely, $\Lambda_A$ introduces freedom in the evolution of phase space fields, which results in the appearance of an asymptotic symmetry. Because the $\mathbf{Z}$ matrix in \eqref{zv.modes} is the identity, the parameter of the associated transformations will also be $r$-independent, the same as the zero modes $v_{A}(u,y)$, based on the formalism presented in Section \ref{Formalism}.
\paragraph{Fall-off of the fields.} 

Using Hamilton's equations, it can be shown  that the standard fall-off of electromagnetic fields \cite{Strominger:2017zoo} also applies in the presence of the $\theta $-term,
\begin{equation}
\begin{array}{llllll}
A_{u} & =\mathcal{O}\left( r^{-1}\right) \,,\quad  & A_{r} & =
\mathcal{O}\left( r^{-2}\right) \,,\quad  & A_{A} & =\mathcal{O}(r^{0})\,,\medskip  \\
\pi^{u} & =0\,, & \pi^{r} & =\mathcal{O}\left( r^{0}\right) \,, &
\pi^{A} & =\mathcal{O}\left( r^{-2}\right) \,,%
\end{array}
\label{fall.EM}
\end{equation}
and that the multipliers behave asymptotically as
\begin{equation}
\lambda _{u} =\mathcal{O}\left( r^{-1}\right) \,,\qquad \lambda _{A} = 
\mathcal{O}\left( r^{0}\right) \,,\qquad A_{u} =\mathcal{O}\left(
r^{0}\right) \,.
\end{equation}

\paragraph{Symmetry transformations.}

The existence of first-class constraints implies that the theory has local symmetries. To analyze them, for each first-class constraint, we define a smeared generator 
\begin{eqnarray}
G[\alpha ] &=&\int \mathrm{d}^{3}x\,\left( \alpha\d_{i}\pi^{i}+\alpha _{u}\pi^{u}\right) ,  \notag \\
S[\eta ] &=&\int \mathrm{d}^{3}x\,\eta _{A}\left( \epsilon \pi^{A}-\sqrt{\gamma}\,\sigma ^{AB}F_{rB}\right) , \label{smear.EM}
\end{eqnarray}
with the local parameters $\alpha(x)$, $\alpha_u(x)$ and  $\eta_A(y)$. 
The above integrals defined in $u=u_0$ are completed by adding corresponding patches of null infinity from the point $u=u_0$ to $u\rightarrow - \epsilon\infty $.

While $\alpha,\alpha_u$ are completely arbitrary parameters that will give a usual $\mathrm{U}(1)$ gauge symmetry, the parameter $\eta$ is only asymptotic because it does not depend on the radial coordinate.  The condition $\partial_r \eta=0$ is necessary to ensure that $S[\eta]$ is the first class variable, because the constraint $\chi^A$ is a combination of first and second class constraint, and the latter ones do not generate symmetries. This can be seen directly from \eqref{zv.modes} and \eqref{Zpar}, because the zero modes $v_A$ and the local parameters $\eta_A$ have the same form. Furthermore, $\eta_A$ is also $u$-independent. Namely,  as we saw in Sec.~\ref{Scalar}, this condition is satisfied for the quantities defined at the null boundary, and for $\eta_A$ also in the bulk because it is $r$-independent.

Indeed,  the generators $G[\alpha ]$ and $S[\eta ]$ are first class. In particular, $S[\eta]$ commute thanks to the $r$-independence of the local parameter,
\begin{eqnarray}
\left\{ S[\eta_1 ],S[\eta_2 ]\right\}  
=-\frac{2\epsilon }{e^{2}}\int \mathrm{d}^{3}x\,\sqrt{\gamma
}\gamma ^{AB}\eta _{1A} \d_{r}\eta_{2B} =0\,,
\end{eqnarray}
while all other PBs are zero directly.

The nontrivial transformations $\delta _{\alpha }=\{\ \ \ ,G[\alpha ]\,\}$ and $\delta
_{\eta }=\{\ \ \ ,S[\eta ]\,\}$ generated by these generators are
\begin{eqnarray}
\delta _{\alpha,\eta }A_{\mu} &=&-\partial _{i}\alpha \,\delta^{i}_{\mu} +\alpha _{u}\, \delta_{\mu}^u +\epsilon \eta _{A} \,\delta_{\mu}^{A}\,,\notag\\
\delta _{\alpha,\eta }\pi ^{r}&=&\sqrt{\gamma }\,\sigma ^{AB}\nabla _{B}\eta _{A}\,.  \label{law.EM}
\end{eqnarray}
 The transformations generated by $\eta _{A}$ are asymptotic, because they
change only the fields that contribute to the boundary,
as given by the fall-off  \eqref{fall.EM}.  Then 
\begin{equation}
\delta _{\eta }A_{(0)A}=\epsilon \eta _{A}\,,\qquad \delta _{\eta }\pi
_{(0)}^{r}=\sqrt{\gamma }\,\sigma ^{AB}\nabla _{B}\eta _{A}\,,
\end{equation}
and they depend on the parameter $\theta $ through the matrix $\sigma ^{AB}$.

Hamilton's equations \eqref{eomP.EM} and \eqref{eomA.EM} are invariant under transformations generated by $G[\alpha ]$, and asymptotically invariant under transformations generated by
$S[\eta ]$, if the multipliers transform as
\begin{equation}
\delta _{\alpha ,\eta }\lambda _{u}=\dot{\alpha}_{u}\,,\qquad \delta
_{\alpha ,\eta }\lambda _{A}=\dot{\eta}_{A}=0\,.
\end{equation}

All equations and laws are consistent with the fall-off of the fields and multipliers when
\begin{equation}
\alpha_u=\mathcal{O}(r^{-1})\,,\qquad \alpha=\mathcal{O}(r^0)\,.
\end{equation} 

\paragraph{Conserved charges.}

The Regge-Teitelboim \cite{Regge:1974zd} conserved charges $Q[\alpha ]$ and $Q_{S}[\eta ]$ are surface terms added to the generators,
\begin{eqnarray}
G_{Q}[\alpha ] &=&G[\alpha ]+Q[\alpha ]\,,  \notag \\
S_{Q}[\eta ] &=&S[\eta ]+Q_{S}[\eta ]\,,
\label{RTG}
\end{eqnarray}
that make them differentiable. This condition determines the variation of the charges
as 
\begin{eqnarray}
\delta Q[\alpha ] &=&-\oint \mathrm{d}^{2}y\,\alpha\delta \pi ^{r}\,,
\notag \\
\delta Q_{S}[\eta ] &=&\oint \mathrm{d}^{2}y\,\sqrt{\gamma }\,\sigma ^{AB}  \eta _{A}\delta A_{B}\,,
\end{eqnarray}
where integrals are evaluated over $\mathbb{S}^2$ at the null infinity $\dI^+_-$ for the future cone and $\dI^-_+$ for the past cone. For the field-independent parameters $\alpha$ and $\eta_{A}$, these charges are integrable, 
\begin{eqnarray}
Q[\alpha ] &=&-\oint \mathrm{d}^{2}y\,\alpha\pi^{r} \,,  \notag \\
Q_{S}[\eta ] &=&\oint \mathrm{d}^{2}y\,\sqrt{\gamma }\,\sigma ^{AB}  \eta _{A}A_{B}\,. \label{charges.EM}
\end{eqnarray}
Because the integrals have to be taken in the limits $r\to \infty$ and $u \to - \epsilon \infty$, the charges depend only on $\pi^r_{(0)}(y)$ and $A_{A(0)}(y)$.  They are also the generators in the reduced phase space because $G_Q \approx Q$ and $S_Q \approx Q_S$.

\paragraph{Symplectic form.}

The next step is to compute the charge algebra. However, setting the constraints strongly to zero in the generators is allowed only in the reduced phase space, where the Poisson brackets have to be replaced by Dirac brackets, and their definition is subtle in our case because of the new first-class constraints which are mixed up with second-class constraints.

A more efficient way to arrive at the charge algebra without introducing the Dirac brackets is using the symplectic formulation of the phase space manifold and differential geometry. Thus, we will first derive the charges \eqref{charges.EM} using
this alternative method, and then we will complete the charge algebra.

The canonical symplectic form reads
\begin{equation}
\omega =\int \diff^{3}x\,\delta \pi ^{\mu }\wedge \delta A_{\mu }\,,
\label{sympl-form}
\end{equation}
where $\delta $ is the exterior derivative on the symplectic manifold. Similarly as previously discussed for smeared generators, the symplectic form \eqref{sympl-form} has to be integrated over a Cauchy surface, such that the integral has to include a patch of null infinity from $u=\mathrm{const}.$~to $u \to \mp \infty$ for the future/past cone. In this way, the obtained charges are indeed evaluated at $\dI^+_-/\dI^-_+$, as argued. We will not write this patch explicitly. For details, see   \cite{Donnelly:2016auv,Geiller:2017xad,Geiller:2017whh,Hosseinzadeh:2018dkh,Speranza:2017gxd}. 
\footnote{Strictly speaking, \eqref{sympl-form} is a presymplectic form 
because it is degenerate due to isometries generated by  \eqref{isometry} with the parameter $\eta_A$
that has no support at $\dI^+_-/\dI^-_+$. Therefore, they have to be quotient out from the covariant configuration space \cite{Lee:1990nz}  
 by requiring that all the parameters are $u$-independent on the null boundary. Namely, it can be shown that the quotient space is the space of time-independent functions. We thank to the anonymous referee for clarifying this issue to us.}

Vector fields $X_{\alpha }$ and $X_{\eta }$ on this manifold are by
definition
\begin{eqnarray}
X_{\alpha } =\int \diff^{4}x\left( \alpha _{u}\,\frac{\delta }{\delta A_{u}}-\partial
_{i}\alpha \,\frac{\delta }{\delta A_{i}}\right) \,,
\end{eqnarray}
and similarly
\begin{eqnarray}
X_{\eta } =\int \diff^{4}x\left( \epsilon \eta _{A}\,\frac{\delta }{\delta A_{A}}+%
\sqrt{\gamma }\sigma ^{AB}\nabla _{B}\eta _{A}\,\frac{\delta }{\delta \pi
^{r}}\right) \,. \label{isometry}
\end{eqnarray}
Such a definition implies that a vector field generates a local transformation
of any phase space function $\mathcal{F} (\pi ,A)$ through the contraction,
$i_{X_{\alpha,\eta}}\delta \mathcal{F} =\delta _{\alpha,\eta }\mathcal{F}$. 

Besides, the above vector fields are isometries of the symplectic form, $\mathscr{L}_{X_{\alpha ,\eta}}\omega =0$. Then, using that the Lie derivative is related to the exterior derivative by
Cartan's formula and that the symplectic form is closed, the above equations are
equivalent to
\begin{equation}
i_{X_{\alpha }}\omega =-\delta Q_{\alpha }\,,\qquad i_{X\eta }\omega
=-\delta Q_{\eta }\,,
\end{equation}%
where $Q_{\alpha }$ and $Q_{\eta }$ are generators. We can show that  $Q_{\alpha }$ and $Q_{\eta }$ exist, and that they coincide with the smeared
generators \eqref{RTG}, namely, $Q_{\alpha }\approx Q[\alpha ]$ and $Q_{\eta
}\approx Q_{S}[\eta ]$.

\paragraph{Charge algebra.}

The algebra of charges, or the algebra of symmetry generators on the reduced phase space, can be obtained as a double contraction of the symplectic form,
\begin{eqnarray}
\left\{ Q[\alpha _{1}],Q[\alpha _{2}]\right\}^{\ast }  &=&-i_{X_{\alpha
_{1}}}i_{X_{\alpha _{2}}}\omega \,,  \notag \\
\left\{ Q_{S}[\eta _{1}],Q_{S}[\eta _{2}]\right\}^{\ast }  &=&-i_{X_{\eta
_{1}}}i_{X_{\eta _{2}}}\omega \,, \\
\left\{ Q[\alpha ],Q_{S}[\eta ]\right\}^{\ast }  &=&-i_{X_{\alpha }}i_{X_{\eta
}}\omega \,.  \notag
\end{eqnarray}
Therefore, we find that the charges $Q[\alpha ]$ and $Q_{S}[\eta ]$ satisfy Abelian
algebra with the central extension,
\begin{eqnarray}
\left\{ Q[\alpha _{1}],Q[\alpha _{2}]\right\}^{\ast }  &=&0\,,  \notag \\
\left\{ Q_{S}[\eta _{1}],Q_{S}[\eta _{2}]\right\}^{\ast }  &=&0\,,  \label{alg.EM} \\
\left\{ Q[\alpha ],Q_{S}[\eta ]\right\}^{\ast }  &=&C[\alpha ,\eta ]\,,  \notag
\end{eqnarray}
where the central charge is
\begin{equation}
C[\alpha ,\eta ]=\oint \mathrm{d}^{2}y\,\sqrt{\gamma }\,\sigma ^{BA}\eta
_{B}\partial _{A}\alpha \,.  \label{C.EM}
\end{equation}%
Notice that the property $\partial _{r}\eta _{A}=0$ of the parameter $\eta
_{A}$ is essential for the closure of the charge algebra. The obtained central extension $C$ is non-trivial since it cannot be removed by redefinition of generators in Abelian theories \cite{deAzcarraga:1995jw}.

Comparing with the central charge \cite{Gonzalez:2023yrz} obtained for $\theta =0$, we conclude that the algebra is not modified by the Pontryagin term, for the given boundary conditions. 
In \cite{Gonzalez:2023yrz}, it is shown explicitly that this algebra can be recognized as a non-semisimple Kac-Moody algebra with a central extension, after expanding the fields in Laurent modes.

\subsection{Electric-magnetic duality}
\label{Duality}

We will show that the obtained asymptotic symmetries and associated charges lead to the electric-magnetic duality in Maxwell electromagnetism with the topological Pontryagin term.

\paragraph{Case $\theta=0$.}

We start with comparing the charges \eqref{charges.EM} with the ones found when $\theta =0$ \cite{Gonzalez:2023yrz}. Then, the Hamiltonian functionals $Q^{0}[\alpha ]$
and $Q_{S}^{0}[\eta ]$ are related to the electric and magnetic charges,
\begin{eqnarray}
Q_{\mathrm{el}}[\alpha ] &\equiv&\oint \mathrm{d}^{2}y\sqrt{\gamma }\,\alpha \,q_{%
\mathrm{el}}=-\frac{r^{2}}{e^{2}}\oint \mathrm{d}^{2}y\sqrt{\gamma }\,\alpha
F_{ur}\,,  \notag \\
Q_{\mathrm{mag}}[\alpha ] &\equiv&\oint \mathrm{d}^{2}y\sqrt{\gamma }\,\alpha
\,q_{\mathrm{mag}}=\frac{1}{2e^{2}}\oint \mathrm{d}^{2}y\,\alpha \,\epsilon
^{AB}F_{AB}\,,  \label{Qel,Qmag}
\end{eqnarray}%
defined as electric and magnetic fluxes of the electromagnetic field through
the 2-sphere at the infinity.\footnote{Again, in all charges, we have to take the limit $u\rightarrow -\epsilon \infty $ to obtain the total flux over the null infinity that enters the conservation law of a radiating system, as explained in Section \ref{Scalar}  (see the text after \eqref{QK}).} The factor $r^2$ comes from $\sqrt{\mathfrak{g}}$ and it is essential to make the electric charge density IR finite. When $\alpha=\mathrm{const.}$, the magnetic charge becomes a topological invariant, and $Q_{\mathrm{mag}}$ reduces to a topological number. 

The relation between $\left( Q^{0},Q_{S}^{0}\right) $ and $\left( Q_{\mathrm{el}},Q_{%
\mathrm{mag}}\right) $ can be obtained by plugging in the Hamilton's
equation $\pi ^{r}=\frac{r^{2}\sqrt{\gamma }}{e^{2}}\,F_{ur}$ in the
functional $Q^{0}[\alpha ]$, which makes it evident that it becomes equal to the electric charge. As regards the functional $Q_{S}^{0}[\eta ]$, first we have to apply the Helmholtz decomposition of the vector field to the divergenceless and rotorless parts,
\begin{equation}
\eta _{A} (\beta,\psi)=\epsilon\nabla _{A}\beta +\sqrt{\gamma }\,\epsilon _{AB}\,\nabla ^{B}\psi
\,,  \label{Helmholtz}
\end{equation}
where $\beta $ and $\psi $ are new $r$-independent parameters that replace $\eta _{A}$, and then we perform the integration by parts to get
\begin{equation}
Q_{S}^{0}[\eta ]=\frac{1}{e^{2}}\oint \mathrm{d}^{2}y\,\left( -\sqrt{\gamma }
\,\epsilon\beta \,\nabla ^{A}A_{A}+\frac{1}{2}\,\psi \,\epsilon ^{AB}F_{AB}\right) .
\end{equation}%
The second term can be recognized as the magnetic charge, while the first one has to be manipulated from Hamilton's equations and the fall-off of the fields,
\begin{equation}
\dot{\pi}^{r}=\frac{1}{e^{2}}\,\sqrt{\gamma }\,\nabla ^{A}\left( \epsilon
F_{uA}-F_{rA}\right) =\left( \frac{\epsilon }{e^{2}}\,\sqrt{\gamma }\,\nabla
_{A}A^{A}\right) ^{. }+\mathcal{O}\left( r^{-1}\right) \,. \label{dotPi}
\end{equation}
Integration in $u$ leads to
\begin{equation}
\pi ^{r}=\frac{\epsilon }{e^{2}}\,\sqrt{\gamma }\,\nabla ^{A}A_{A}+k(y)+\mathcal{O}\left( r^{-1}\right) \, \label{integrated}
\end{equation}
up to an arbitrary angle-dependent function which we set to zero for simplicity, $k(y)=0$. Since $\pi ^{r}=\frac{r^{2}\sqrt{\gamma }}{e^{2}}\,F_{ur}$, we get
\begin{equation}
\nabla ^{A}A_{A}=\epsilon r^{2}F_{ur}+\mathcal{O}\left( r^{-1}\right) \,,
\label{divA}
\end{equation}
recognizing the electric charge density. In sum, the Hamiltonian charges, when $\theta =0$, are identified with the Lagrangian quantities through the mapping
\begin{eqnarray}
Q^{0}[\alpha ] &\equiv&Q_{\mathrm{el}}[\alpha ]\,,  \notag \\
Q_{S}^{0}[\eta ] &\equiv&Q_{\mathrm{el}}[\beta]+Q_{\mathrm{mag}}[\psi ]\,, \label{Q0}
\end{eqnarray}
 which means that $Q^{0}$ is just the standard electric charge, while $Q_{S}^{0}$ is the magnetic charge up to an additional electric term.

To find whether the electric and magnetic charges commute, we express the known algebra \eqref{alg.EM} in terms of $Q_{\mathrm{el}}$ and $Q_{\mathrm{mag}}$ given by  \eqref{Q0}. We get
\begin{eqnarray}
\left\{ Q_{\mathrm{el}}[\alpha _{1}],Q_{\mathrm{el}}[\alpha _{2}]\right\}^{\ast }
&=&0\,,  \notag \\
\left\{ Q_{\mathrm{el}}[\alpha ],Q_{\mathrm{mag}}[\psi ]\right\}^{\ast }
&=&C^0[\alpha ,\eta (\beta ,\psi )]\,,\label{alg_aux}\\
\left\{ Q_{\mathrm{mag}}[\psi _{1}],Q_{\mathrm{mag}}[\psi _{2}]\right\}^{\ast }
&=&C^0[\beta _{2},\eta (\beta _{1},\psi _{1})]-C^0[\beta _{1},\eta (\beta
_{2},\psi _{2})]\,,   \notag
\end{eqnarray}
where the central charge when $\theta=0$ has the form
\begin{equation}
C^0[\alpha ,\eta (\beta ,\psi )]=\frac{1}{e^{2}}\oint \mathrm{d}^{2}y\,\epsilon \sqrt{\gamma }\,\nabla ^{A}\alpha \,\nabla _{A}\beta = C[\alpha ,\eta (\beta ,0)]  \equiv
C_{\mathrm{EM}}[\alpha ,\beta ]\,. \label{cEM}
\end{equation}
The $\psi $-term does not contribute to \eqref{cEM} when the sphere is smooth because
$\frac{1}{e^{2}}\oint \mathrm{d}^{2}y\,\epsilon ^{AB}\partial _{A}\alpha
\partial _{B}\psi =\frac{1}{e^{2}}\oint \mathrm{d}^{2}y\,\partial _{A}\left(
\epsilon ^{AB}\alpha \partial _{B}\psi \right) =0$. In addition, the electric-magnetic central extension is symmetric, $C_{\mathrm{EM}}[\alpha ,\beta ]=C_{\mathrm{EM}}[\beta ,\alpha ]$. Hence, the algebra becomes
\begin{eqnarray}
\left\{ Q_{\mathrm{el}}[\alpha ],Q_{\mathrm{mag}}[\psi ]\right\}^*  =C_{\mathrm{EM}}[\alpha ,\psi ]\,,  \label{QeQm0}
\end{eqnarray}
and all other PBs are zero. 

It is worth emphasizing that the mapping \eqref{Q0} that identifies the
electric and magnetic fluxes with the Hamiltonian charges is not invertible
when $\alpha =0$ or $\psi =0$, producing ambiguities in the definition of charges. For instance, we could have $Q_{\mathrm{el}}[\alpha ]=Q^{0}[\alpha ]$ or $Q'_{\mathrm{el}}[\alpha]=Q_{S}^{0}[\eta (\alpha ,\psi =0)]$ for the electric flux. The
corresponding Abelian algebras \eqref{alg_aux} would have different
central extensions. Thus, we have to require $\alpha, \psi \neq 0$ to have invertible  mapping $(Q_{\mathrm{el}},Q_{\mathrm{mag}})\leftrightarrow (Q^0,Q^0_S)$. Then the $\mathrm{U}(1)$ charge is given by a usual expression in both Hamiltonian and Lagrangian formalisms.

In \cite{Hosseinzadeh:2018dkh,Freidel:2018fsk}, the authors also discuss the electric-magnetic duality in electromagnetism and the corresponding charge algebra, finding the central charge $C_{\mathrm{EM}}\propto \int \mathrm{d}^{2}y\,\epsilon ^{AB}\partial _{A}\psi \partial _{B}\alpha = \oint \psi \mathrm{d}\alpha$.  They obtain only the $\eta (0,\psi )$-term of the central charge, which vanishes in our case, because we choose configurations with the smooth boundary of the 2-sphere, leaving only the  $\eta (\beta,0 )$-term. In contrast, in \cite{Freidel:2018fsk}, the authors look at non-trivial configurations with non-smooth boundaries, such as the point-like electric charge that plays the role of a monopole for a dual potential. The electric-magnetic duality in the context of soft charges has also been discussed in \cite{Hosseinzadeh:2018dkh}.

One of the important consequences of the fact that $C_{\mathrm{EM}}\neq 0$ is that the quantized electric and magnetic charges cannot be measured simultaneously, according to Heisenberg's uncertainty principle. This fact could explain why a magnetic monopole has not been observed yet \cite{Freidel:2018fsk}.

\paragraph{Case $\theta \neq 0$.}

We can repeat the previous computation when $\theta \neq 0$. First, we note that, using the definition of the radial momentum \eqref{pi.EM},
the charge $Q$ becomes a linear combination of both electric and magnetic
terms. On the other hand, we also have
\begin{equation}
\dot{\pi}^{r}=\sqrt{\gamma }\,\nabla _{A}\left( \epsilon F_{uB}\sigma ^{BA}-\frac{1}{e^{2}}\,\gamma ^{AB}F_{rB}\right) \,,
\end{equation}
which in the asymptotic expansion leads to\footnote{Similarly as in \eqref{integrated}, we neglect the integration constant, $k(y)=0$.}
\begin{equation}
\dot{\pi}^{r}=\left( \epsilon \sqrt{\gamma }\,\sigma ^{BA}\,\nabla
_{A}A_{B}\right) ^{\cdot }+\mathcal{O}\left( r^{-1}\right) \quad \Rightarrow
\quad \pi ^{r}=\epsilon \sqrt{\gamma }\sigma ^{BA}\nabla _{A}A_{B}+\mathcal{O}(r^{-1})\,, \label{integrated-theta}
\end{equation}
such that, replacing $\pi ^{r}$ from the Hamilton's equation, we find that
the $\theta $-term does not affect the identity (\ref{divA}). Hence, we get the identifications that map Lagrangian to Hamiltonian charges,
\begin{eqnarray}
Q[\alpha ] &\equiv&Q_{\mathrm{el}}[\alpha ]-Q_{\mathrm{mag}}[\theta\alpha
]\,,  \notag \\
Q_{S}[\eta ] &\equiv&Q_{\mathrm{el}}[\beta -\theta\,\psi ]+Q_{\mathrm{mag}}[\psi +\theta\beta]\,. \label{Q,em,mag}
\end{eqnarray}
We conclude that, when $\theta \neq 0$, the electric charge acquires a magnetic part and the
magnetic charge acquires an electric one. Thus, it is more useful to define the total  charge,
\begin{equation}
Q_{\mathrm{tot}}[\xi ]=Q[\alpha ]+Q_{S}[\eta ]\equiv\oint\limits_{\mathbb{S}_{\infty }^{2}}\mathrm{d}^{2}y\,\sqrt{\gamma }\,\xi ^{T}q\,, \label{tot}
\end{equation}
where now both electric and magnetic parameters $\xi$, associated with the charges $q$ given by \eqref{Qel,Qmag}, depend on three parameters $\left( \alpha ,\beta ,\psi \right) $,
\begin{equation}
\xi =\left(
\begin{array}{c}
\xi _{\mathrm{el}} \\
\xi _{\mathrm{mag}}%
\end{array}%
\right) =\left(
\begin{array}{c}
\alpha +\beta -\theta \psi  \\
\psi -\theta \alpha +\theta \beta
\end{array}%
\right) \,,\qquad q=\left(
\begin{array}{c}
q_{\mathrm{el}} \\
q_{\mathrm{mag}}%
\end{array}%
\right) \,.  \label{el,mag}
\end{equation}
Because there are two charges and three parameters, only two of them, say $\xi _{\mathrm{el}}$ and $\xi _{\mathrm{mag}}$, generate independent field transformations. An arbitrariness in the choice of these parameters (and therefore generators) leads to the electric-magnetic duality. 

Namely, the total charge is $\mathrm{SO}(2)$ invariant because the quadratic form $\xi ^{T}q$ is unchanged under orthogonal transformations.
Explicitly, an invertible $\mathrm{SO}(2)$ transformation $\xi \rightarrow
\xi ^{\prime }$ that performs a rotation between asymptotic parameters and
charges with the constant parameter $\mu $ is given by
\begin{equation}
\left(
\begin{array}{c}
\xi' _{\mathrm{el}} \\
\xi' _{\mathrm{mag}}
\end{array}
\right) =\left(
\begin{array}{cc}
\cos \mu  & \sin \mu  \\
-\sin \mu  & \cos \mu
\end{array}
\right) \left(
\begin{array}{c}
\xi _{\mathrm{el}} \\
\xi _{\mathrm{mag}}
\end{array}
\right) \,.
\end{equation}

Because there are three parameters $\left( \alpha ,\beta ,\psi \right) $ and only two charges in $q$, a more general invertible law $(\alpha ,\beta ,\psi) \to (\alpha' ,\beta' ,\psi')$ is given by the $\mathrm{GL}(2,\mathbb{R})$ transformations. 
Indeed, requiring the invariance of $\xi^T q$ in terms of these three parameters, the most general change is
\begin{eqnarray}
\alpha' &=&\left( \frac{1+\theta ^{2}}{2\theta }\,a+\frac{1-\theta
^{2}}{2\theta }\,\sin \mu +\cos \mu \right) \alpha +\frac{1+\theta ^{2}}{2\theta }\,\left( b+\sin \mu \right) \beta +\frac{1+\theta ^{2}}{2\theta }
\,\left( c-\cos \mu \right) \psi \,,  \notag \\
\beta ^{\prime } &=&\left( -\frac{1-\theta ^{2}}{2\theta }\,a-\frac{1+\theta
^{2}}{2\theta }\,\sin \mu \right) \alpha +\left( -\frac{1-\theta ^{2}}{%
2\theta }\,\left( b+\sin \mu \right) +\cos \mu \right) \beta   \notag \\
&&+\left( \frac{1-\theta ^{2}}{2\theta }\,\left( -c+\cos \mu \right) +\sin
\mu \right) \psi \,, \\
\psi ^{\prime } &=&a\alpha +b\beta +c\psi \,,  \notag
\end{eqnarray}
where $a$, $b$, $c$ and $\mu$ are real numbers. The transformation exists when $\theta \neq 0$; thus, the electric-magnetic duality is realized only in the presence of the $\theta $-term. 
It is standard to normalize a global factor of these parameters such that the transformation becomes $\mathrm{SL}(2,\mathbb{R})$.

The described electromagnetic duality is a direct consequence of the new asymptotic symmetry generated by zero modes of the symplectic matrix.

Similarly, as in the case $\theta =0$, we can investigate the algebra of $Q_{\mathrm{el}}$ and $Q_{\mathrm{mag}}$ to analyze how the central charge \eqref{C.EM} will translate under duality transformations. Recall that it can be written in terms of the three parameters
\begin{equation}
C[\alpha ,\eta (\beta ,\psi )]=\frac{\epsilon }{e^{2}}\oint \mathrm{d}^{2}y\, \sqrt{\gamma }\,\partial _{A}\alpha \nabla ^{A}\left( \beta -\theta \psi \right) \,,  \label{C.SEM}
\end{equation}
where we neglected the boundary terms taken over the  2-sphere at the null infinity.

To this end, we have to invert eqs.~\eqref{Q,em,mag} and express the electric and magnetic charges in terms of $Q$ and $Q_{S}$, which is ambiguous, since there are more parameters than charges. Thanks to $\mathrm{SL}(2,\mathbb{R})$ symmetry, one parameter can always be set to zero, except $\alpha $, otherwise the relations \eqref{Q,em,mag}  are not invertible.  
To choose which parameter to set to zero, we remember from the $\theta =0$ case given by \eqref{Q0} that $\alpha $ and $\beta $ have an electric nature and $\psi $ magnetic nature. The introduction of $\theta \neq 0$ mixes them up and they all contribute to both electric and magnetic terms, but it is more natural to set $\beta =0$ to be able to study $\theta \rightarrow 0$ limit at the end. Then, from \eqref{el,mag}, we can express $\alpha $ and $\psi $ in terms of the true electric and magnetic gauge parameters $\xi _{\mathrm{el}}$ and $\xi _{\mathrm{mag}}$ as
\begin{equation}
\alpha =\frac{\xi _{\mathrm{el}}+\theta \xi _{\mathrm{mag}}}{1-\theta ^{2}} \,,\qquad \psi =\frac{\xi _{\mathrm{mag}}+\theta \xi _{\mathrm{el}}}{%
1-\theta ^{2}}\,,  \label{el-mag}
\end{equation}%
seeing that indeed $\alpha $ has origin in the electric and $\psi $ in the
magnetic part. We assume $\theta \neq \pm 1$. The (anti)self-dual points $\theta =\pm 1$ could change the asymptotic behavior of the fields, so they have to be discussed separately.

When $\beta =0$, the inverted relations \eqref{Q,em,mag} have the form
\begin{eqnarray}
Q_{\mathrm{el}}[\alpha ] &=&\frac{1}{1-\theta ^{2}}\,\left(
\rule[2pt]{0pt}{12pt}Q[\alpha ]+\theta Q_{S}[\eta (0,\alpha )]\right) \,,
\notag \\
Q_{\mathrm{mag}}[\psi ] &=&\frac{1}{1-\theta ^{2}}\,\left(
\rule[2pt]{0pt}{12pt}\theta Q[\psi ]+Q_{S}[\eta (0,\psi )]\right) \,.
\end{eqnarray}
From the algebra \eqref{alg.EM} of $Q$ and $Q_{S}$, it is straightforward to compute the PBs of $Q_{\mathrm{el}}$ and $Q_{\mathrm{mag}}$. Using the property $C[\alpha ,\eta (0,\psi )]=C[\psi ,\eta (0,\alpha )]$, we arrive at the result
\begin{eqnarray}
\left\{ Q_{\mathrm{el}}[\alpha _{1}],Q_{\mathrm{el}}[\alpha _{2}]\right\}^*
&=&0\,,  \notag \\
\left\{ Q_{\mathrm{el}}[\alpha ],Q_{\mathrm{mag}}[\psi ]\right\}^*  &=&C_{\mathrm{EM}}[\alpha ,\psi ]\,, \label{algebraTheta}\\
\left\{ Q_{\mathrm{mag}}[\psi _{1}],Q_{\mathrm{mag}}[\psi _{2}]\right\}^*
&=&0\,.  \notag
\end{eqnarray}
There is only one central extension,
\begin{equation}
C_{\mathrm{EM}}[\alpha,\psi]=\frac{1}{1-\theta ^{2}}\,C[\alpha
,\eta (0,\psi)]=-\frac{\theta }{1-\theta ^{2}}\,\frac{\epsilon }{e^{2}}\oint \mathrm{d}^{2}y\,\sqrt{\gamma }\,\partial _{A}\alpha \nabla
^{A}\psi\,. \label{C_EM}
\end{equation}
The result depends on  $\theta $ and, in the limit $\theta \rightarrow 0$, it reproduces the old result $C^{0}$ given by \eqref{cEM}, where $\psi'=-\theta \psi $ should be kept finite in this limit, as this parameter has the origin in the effective parameter $\psi'=\beta -\theta \psi $ of the central charge \eqref{C.SEM}, after setting $\beta =0$.

It is worth emphasizing that, although the total charge \eqref{tot} remains unchanged for different choices of two independent parameters, this is not necessarily true for the corresponding algebras. It is therefore not clear whether the various Abelian algebras presented in this section would have physically equivalent central extensions. 

\section{Electromagnetism coupled to the scalar field}
\label{EM with scalar}

Let us consider the full interacting theory with the massless scalar field $\phi(x)$ and the electromagnetic field $A_\mu(x)$, coupled through the Pontryagin density as given in the action \eqref{s.EM}.
Compared to the previous section and the action \eqref{Maxwell}, we made the replacement $\theta \to \theta \phi (x)$, making the coupling parameter dynamic.  This changes the fall-off of the interaction from $\theta=\mathcal{O}(r^0)$ to  $\theta \phi = \mathcal{O}(r^{-1})$.

Note that the action \eqref{s.EM} possesses a global shift symmetry $\phi(x) \to \phi(x)+ const$ because the interaction term becomes a topological invariant for a constant scalar.

To pass to the Hamiltonian description of the system, we introduce the momenta $(\pi,\pi^\mu)$ canonically conjugated to the fields $(\phi,A_\mu)$ as
\begin{equation}
\begin{array}{llll}
\pi  & =\epsilon r^{2}\sqrt{\gamma }\partial _{r}\phi \,,\qquad  & \pi^{r} & =\dfrac{r^{2}}{e^{2}}\,\sqrt{\gamma }F_{ur}+\dfrac{\theta}{2e^2}\,
\phi \,\epsilon ^{AB}F_{AB}\,,\medskip  \\
\pi^{u} & =0\,, & \pi^{A} & =\epsilon \sqrt{\gamma }\,\sigma^{AB} F_{rB}\,, \label{pi}
\end{array}
\end{equation}
where now the matrix $\sigma^{AB}$ is not covariantly constant because it depends on the scalar field,
\begin{equation}
\sigma ^{AB}(\phi )=\frac{1}{e^{2}}\left(\gamma ^{AB}-\frac{
\epsilon \theta }{\sqrt{\gamma }}\,\phi \,\epsilon ^{AB}\right)\,.
\label{sigma(phi)}
\end{equation}
The canonical PBs are
\begin{equation}
\left\{ \phi ,\pi'\right\} =\delta ^{(3)}(x-x')\,,\qquad
\left\{ A_{\mu },\pi ^{\nu \prime }\right\} =\delta _{\mu }^{\nu }\,\delta ^{(3)}(x-x')\,,
\end{equation}
where all other brackets are zero.\medskip

After the Legendre transformation of the Lagrangian density, and neglecting the boundary terms, the total Hamiltonian can be cast into the form
\begin{equation}
\mathcal{H}_{T}=\mathcal{H}_{0}-A_{u}\chi +\lambda \hat\chi+\lambda
_{A}\chi^{A}+\lambda _{u}\pi^{u}\,.
\end{equation}
Based on \eqref{pi}, we introduced the primary constraints,
\begin{eqnarray}
\pi^{u} &\approx &0\,,
\notag \\
\hat\chi  &=&\epsilon \pi -r^{2}\sqrt{\gamma }\,\partial _{r}\phi \approx 0\,, \label{new.con} \\
\chi^{A} &= &\epsilon \pi^{A}-\sqrt{\gamma }\,\sigma
^{AB}F_{rB}\approx 0\,,  \notag
\end{eqnarray}
and the associated multipliers $\lambda(x)$, $\lambda_A(x)$ and $\lambda_u(x)$. We also denoted
\begin{eqnarray}
\mathcal{H}_{0} &=&\frac{\pi ^{2}}{2r^{2}\sqrt{\gamma }}+\frac{1}{2}\sqrt{\gamma }\,\nabla^{A}\phi \nabla_{A}\phi +\frac{e^{2}}{2r^{2}\sqrt{\gamma }}\,\pi^{r}\left( \pi ^{r}-\frac{\theta}{e^2}\,\phi
\,\epsilon ^{AB}F_{AB}\right)   \notag \\
&&+\frac{e^2\pi^{A}\tilde{\pi}_{A}}{2(1+\theta^2\phi^2)\sqrt{\gamma }}
+\frac{\sqrt{\gamma }}{4e^2r^{2}}\,(1+\theta^2\phi^2)F_{AB}\tilde{F}^{AB}\,,
\end{eqnarray}
and recognized that the temporal component of the gauge field $A_u$ plays the role of the Hamiltonian multiplier of the secondary constraint
\begin{equation}
\dot{\pi}^{u}=0\quad \Rightarrow \quad \chi =\partial_{i}\pi^{i}\approx 0\,. \label{sec}
\end{equation}
This secondary constraint has vanishing PBs with all the primary constraints  \eqref{new.con}.

So far, everything is as expected -- we have two new primary constraints $\hat{\chi}$ and $\chi^A$, one for the scalar field and another for the electromagnetic field, that are the consequence of working in the null foliation of spacetime. It remains to analyze their evolution, as well as the secondary constraint, $\chi$. This is because we have to ensure that all constraints will remain on the constraint surface in time.

It is useful first to find Hamilton's equations. For the gauge field momenta, they read  
\begin{eqnarray}
\dot{\pi}^{u} &=&\chi\approx 0\,, \notag \\
\dot{\pi}^{r} &=&\sqrt{\gamma}\,\nabla_{A}\left(\lambda_{B}\sigma ^{BA}\right) \,,   \\
\dot{\pi}^{A} &=& \frac{1}{r^{2}}\,\partial_{B}\left( \frac{\theta\phi }{\sqrt{\gamma }}\,\epsilon^{AB}\pi^{r}-\frac{1+\theta^2\phi^2}{e^2}\, \sqrt{\gamma }\,\tilde{F}^{AB}\right) -\partial_{r}\left(\sqrt{\gamma}\,\sigma ^{BA}\lambda _{B}\right)\notag \,,
\end{eqnarray}
while the equations for the gauge fields read 
\begin{eqnarray}
\dot{A}_{u} &=&\lambda _{u}\,,  \notag \\
\dot{A}_{r}&=&\frac{e^{2}}{r^{2}\sqrt{\gamma }}\,\left(
\pi ^{r}-\frac{\theta}{2e^2} \,\phi \,\epsilon ^{AB}F_{AB}\right) +\partial_{r}A_{u}\,,  \notag \\
\dot{A}_{A} &=&
\frac{e^2\tilde{\pi}_{A}}{(1+\theta^2\phi^2) \sqrt{\gamma}}+\partial_{A}A_{u}+\epsilon \lambda _{A}\,. \label{Ham.1}
\end{eqnarray}
Furthermore, for the scalar field and scalar momentum, the equations are
\begin{eqnarray}
\dot{\phi} &=&\frac{\pi }{r^{2}\sqrt{\gamma }}+\epsilon \lambda \,,  \notag 
\\
\dot{\pi} &=&\sqrt{\gamma }\,\tilde{\square}\phi +\frac{\theta }{2r^{2}\sqrt{\gamma }}\,\pi^{r}\epsilon ^{AB}F_{AB}-\theta^{2}\frac{\sqrt{\gamma }}{2e^2r^{2}}\,\phi F_{AB}\tilde{F}^{AB}  \notag \\
&&+\frac{e^{2}\theta^{2}\pi^{A}\tilde{\pi}_{A}\phi }{\sqrt{\gamma }(1+\theta^2\phi^2)^{2}}-\sqrt{\gamma }\partial _{r}(\lambda
r^{2})- \frac{\epsilon\theta}{e^2}\,
\epsilon^{AB}\lambda_{A}\,F_{rB}\,. \label{Ham.2}
\end{eqnarray}
These equations of motion are consistent with the standard fall-off of the fields proposed in \eqref{fall.S} and \eqref{fall.EM},
\begin{equation}
\begin{array}{llllllll}
A_{u} & =\mathcal{O}\left( r^{-1}\right) \,,\quad  & A_{r} & =
\mathcal{O}\left( r^{-2}\right) \,,\quad  & A_{A} & =\mathcal{O}(r^{0})\,, \quad &\phi & =\mathcal{O}\left(r^{-1}\right)\,, \medskip  \\
\pi^{u} & =0\,, & \pi^{r} & =\mathcal{O}\left( r^{0}\right) \,, &
\pi^{A} & =\mathcal{O}\left( r^{-2}\right) \,, &\pi& =\mathcal{O}\left(r^{0}\right)\,,
\end{array} \label{falloff2}
\end{equation}
where the multipliers behave as
\begin{equation}
\lambda _{u}=\mathcal{O}\left( r^{-1}\right) \,,\qquad \lambda _{A}=
\mathcal{O}\left( r^{0}\right) \,,\qquad \lambda=\mathcal{O}\left(
r^{-1}\right) \,.
\end{equation}

Let us turn now to the discussion of the evolution of the obtained constraints.

\paragraph{Consistency conditions.}

We already found one secondary constraint \eqref{sec}. Before we proceed, let us compute the non-vanishing brackets between the primary and secondary constraints. We get
\begin{eqnarray}
\{\hat\chi ,\hat\chi'\} &=&-2\epsilon r\sqrt{\gamma }\,\left( 1+r\partial
_{r}\right) \delta ^{(3)}\,,  \notag \\
\{\hat\chi ,\chi'^A\} &=&-\frac{\theta}{e^2} \,\epsilon^{AB}F_{rB}\,\delta^{(3)}\,,   \label{PBs}\\
\left\{ \chi ^{A},\chi^{\prime B}\right\}  &=&\Omega^{AB}(x,x')\,, \notag
\end{eqnarray}
where the symplectic matrix depends on the derivative of the scalar field, 
\begin{equation}
\Omega^{AB}(x,x')=-\frac{2\epsilon }{e^{2}}\,
\sqrt{\gamma }\gamma ^{AB}\partial _{r}\delta^{(3)}-\frac{\theta}{e^2} \,\epsilon^{AB}\partial_r\phi\,\delta
^{(3)} \,. \label{OmegaYMS}
\end{equation}
Note that the symplectic matrix (and the zero modes) now depends on the coupling constant $\theta$, due to the interaction with the scalar. This is different from the case discussed in Section \ref{EM+P}, where the Pontryagin term was topological.

As regards the consistency condition of the constraint $\chi$, it is satisfied automatically. For the other two constraints, $\chi _{s}=(\hat\chi ,\chi^{A})$, the result depends on the rank of the matrix $\{\chi _{s}(x),\chi_{s'}(x') \}$ corresponding to the brackets \eqref{PBs}. We also know that the constraints $\chi _{s}$ have vanishing brackets with all other constraints,
\begin{equation}
\left\{ \chi _{s},\chi'\right\} =0\,,\qquad \left\{ \chi _{s},\pi'^u\right\} =0\,.
\end{equation}

Thus, let us define the symplectic matrix \eqref{matrix},
\begin{equation}
\{\chi _{s},\chi' _{s'}\}=\Omega_{ss'}(x,x')=\left(
\begin{array}{cc}
-2\epsilon r\sqrt{\gamma }\,\left( 1+r\partial _{r}\right) \delta ^{(3)} & -\frac{\theta}{e^{2}}\,\epsilon ^{BD}F_{rD}\,\delta ^{(3)} \\
\frac{\theta}{e^{2}}\,\epsilon ^{AC}F_{rC}\,\delta ^{(3)} & \Omega
^{AB}(x,x')
\end{array}
\right) \,.  \label{hatO}
\end{equation}
Then, requiring that the constraints $\chi _{s}\approx 0$ remain on the constraints surface during their evolution, we arrive at the equation for the multipliers of the form   \eqref{eq.lambda}, 
\begin{equation}
\dot{\chi}_{s}=\int \mathrm{d}^{3}x'\,\left\{\chi _{s},\mathcal{H}'
_{T}\right\} = -Y_s+\int \mathrm{d}^{3}x'\,\Omega_{ss'}(x,x')\lambda
^{\prime s'}=0\,,  \label{psi dot}
\end{equation}%
where $\lambda ^{s}=\left( \lambda ,\lambda _{A}\right) $ are the  multipliers, and the quantity $Y_s=(Y,Y^A)$ is given by 
\begin{equation}
Y =\int \mathrm{d}^{3}x'\,\{\mathcal{H}'_{0},\chi \}\,,
\qquad
Y^{A} =\int \mathrm{d}^{3}x'\{\mathcal{H}'_{0},\chi
^{A}\}\,.
\end{equation}
Its explicit form can be obtained using Hamilton's equations \eqref{Ham.1}
and \eqref{Ham.2}, 
\begin{eqnarray}
Y &=&-\epsilon \sqrt{\gamma }\,\tilde{\square}\phi +r^{2}\,\partial
_{r}\left( \frac{\pi }{r^{2}}\right) -\frac{\epsilon \theta\pi ^{r}}{%
2r^{2}\sqrt{\gamma }}\,\epsilon ^{AB}F_{AB}+\frac{\epsilon \theta ^{2}\sqrt{%
\gamma }}{2e^{2}r^{2}}\,\phi F_{AB}\tilde{F}^{AB}-\frac{\epsilon
\,e^{2}\theta ^{2}\pi ^{A}\tilde{\pi}_{A}\phi }{\sqrt{\gamma }\left(
1+\theta ^{2}\phi ^{2}\right) ^{2}}\,,  \notag \\
Y^{A} &=&\frac{\epsilon }{r^{2}}\,\partial _{B}\left( -\frac{\theta \phi }{%
\sqrt{\gamma }}\,\epsilon ^{AB}\pi ^{r}+\frac{1+\theta ^{2}\phi ^{2}}{e^{2}}%
\,\sqrt{\gamma }\tilde{F}^{AB}\right) -\frac{\epsilon \theta \pi }{r^{2}e^{2}%
\sqrt{\gamma }}\,\epsilon ^{AB}F_{rB} \\
&&+e^{2}\,\sigma ^{AB}\partial _{r}\left( \frac{\tilde{\pi}_{B}}{1+\theta
^{2}\phi ^{2}}\right) +\frac{\sqrt{\gamma }}{r^{2}}\,\sigma ^{AB}\partial
_{B}\left( \frac{\frac{\theta }{2}\phi \,\epsilon ^{AB}F_{AB}-e^{2}\pi ^{r}}{%
\sqrt{\gamma }}\right) \,.  \notag
\end{eqnarray}

The equations \eqref{psi dot} determine the multipliers $\lambda^s$. As discussed before, when the symplectic matrix has zero modes, there are indefinite multipliers and, therefore, there exist first-class constraints and associated symmetry generators. It means that we have to study zero modes of $\Omega$ if we want to identify the symmetry generators.

\paragraph{First-class constraints.}

Let us apply the method described in Section \ref{Formalism} for the identification of symmetry generators \eqref{smeared} or \eqref{Zpar} directly from the symplectic matrix \eqref{hatO}. The zero mode
equation \eqref{z.modes} reads
\begin{equation}
\int \mathrm{d}^{3}x'\,\Omega_{ss'}(x,x')V^{\prime s'}=0\,,\qquad V^{s}=\left(
\begin{array}{c}
V \\
V_{A}
\end{array}
\right) ,  \label{zeromode}
\end{equation}
or equivalently,
\begin{eqnarray}
0 &=&\partial _{r}(rV)+\frac{\epsilon \theta }{2e^{2}r\sqrt{\gamma }}%
\,\epsilon ^{AB}F_{rB}\,V_{A}\,,  \notag \\
0 &=&\partial _{r}V^{A}-\frac{\epsilon \theta }{2\sqrt{\gamma }}\,\epsilon
^{AB}F_{rB}\,V + \frac{\epsilon \theta }{2\sqrt{\gamma }}
\,\epsilon ^{AB}\partial _{r}\phi \,V_{B}\,.
\label{PDE}
\end{eqnarray}
This is a coupled system of partial differential equations  
of the form
\begin{equation}
\partial _{r}\left(
\begin{array}{c}
rV \\
V^{A}%
\end{array}%
\right) +\mathbf{M}(x)\left(
\begin{array}{c}
rV \\
V^{A}%
\end{array}%
\right) =0\,,
\end{equation}%
where the matrix $\mathbf{M}(x)=\mathbf{M}(\phi (x),A_\mu (x))$ is known.
Imposing the boundary conditions 
\be
\left( rV,V^{A}\right) \rightarrow \left(
v(u,y),v^{A}(u,y)\right)\,, \qquad r\rightarrow \infty\,, 
\ee 
the formal general solution
always exists as a path-ordered exponential, 
\begin{equation}
\left(
\begin{array}{c}
rV \\
V^{A}%
\end{array}%
\right) =\dP\mathrm{e}^{\int^{\infty}_r\mathrm{d}r'\,\mathbf{M}(u,r',y)}\left(
\begin{array}{c}
v \\
v^{A}
\end{array}
\right) \,,
\end{equation}
and it is highly non-local because it involves the non-local operators $\int^{\infty}_r \mathrm{d}r'\,\mathbf{M}(u,r',y)$.  To evaluate it, it is
more insightful to apply some approximation. One possibility is the large-$r$ approximation but, in our case, it is not applicable because the zero modes generally exist in the whole spacetime. 

Therefore, we will work in the weak-coupling approximation and assume that $\theta $ is a small interaction
parameter. Separating the dependence in the parameter in $\mathbf{M}=\mathbf{M}_{0}+\theta \mathbf{M}_{1}$, which satisfy $\left[
\mathbf{M}_{0},\mathbf{M}_{1}\right] \neq 0$, the above exponential has to
be expanded using the  Baker-Campbell-Hausdorff formula. 

However, it is simpler and more intuitive to solve the equations for given boundary conditions using the method of successive approximations.
To this end, we seek for a solution in the form
\begin{equation}
V^{s}=V_{0}^{s}+\theta V_{1}^{s}+\theta ^{2}V_{2}^{s}+\mathcal{O}(\theta ^{3})\,,
\end{equation}
and obtain a coupled system of linear differential equations equations for the components $V^s_n$,
\begin{equation}
\begin{array}{ll}
n=0:\quad\medskip  & \partial _{r}\left( rV_{0}\right) =0\,,\qquad \partial
_{r}V_{0}^{A}=0\,, \\
n\geq 1:\medskip  & \partial _{r}\left( rV_{n}\right) =-\dfrac{\epsilon }{2e^{2}}\,\dfrac{1}{r\sqrt{\gamma }}\epsilon ^{AB}F_{rB}\,V_{n-1A}\,, \\
\medskip  & \partial _{r}V_{n}^{A}=\dfrac{\epsilon }{2\sqrt{\gamma }}
\,\epsilon ^{AB}\left( F_{rB}\,V_{n-1}-\partial _{r}\phi \,V_{n-1B}\right)
\,.
\end{array}
\label{recursive}
\end{equation}
These equations are solved in Appendix \ref{Zmodes} up to the third order in $\theta$. They explicitly depend on the  non-local functions with the origin in the non-local operators $\int^{\infty}_r \mathrm{d}r'\,\mathbf{M}(u,r',y)$,
\begin{equation}
\begin{array}{llll}
\Delta ^{A} & =-\dfrac{1}{\sqrt{\gamma }}\bigintss\limits^{\infty}_r\dfrac{\mathrm{d}
r'}{r'}\,\epsilon ^{AB}F'_{rB}\,,
\medskip  & \hat{\Delta}_{A} & =-\bigintss\limits^{\infty}_r\dfrac{\mathrm{d}r'
}{r'}\,\phi'F'_{rA}\,, \\
\Delta  & =-\dfrac{1}{\sqrt{\gamma }}\,\epsilon ^{AB}\bigintss\limits^{\infty}_r\dfrac{\mathrm{d}r'}{r'}\,\Delta' _{A}F'_{rB}\,,\quad
& \Delta ^{AB} & =-\dfrac{1}{\sqrt{\gamma }}\,\epsilon ^{AC}\bigintss\limits^{\infty}_r%
\dfrac{\mathrm{d}r'}{r'}\,F'_{rC}\,\Delta^{\prime B}\,, \label{nonlocal}
\end{array}
\end{equation}
where $\mathcal{F}' = \mathcal{F}(u,r',y)$ for the fields under the integrals. 

We obtain that the solution $V^s$ depends on arbitrary functions $v^{s}=v_{0}^{s}+\theta v_{1}^{s}+\theta ^{2}v_{2}^{s}+\mathcal{O}(\theta ^{3})$ that depend only on the coordinates $(u,y^A)$.  Since the functions $v^{s}$ are just a re-summation of the modes at different orders of $\theta$, the same result would have been obtained only with $v^s_0$. Thus, from now on, we take that $v_s$ is independent of $\theta$ for the sake of simplicity.

There are two coupled asymptotic zero modes $v^s=(v,v^A$), where $v(u,y)$ is associated with the scalar field, and $v^{A}(u,y)$ is associated with the electromagnetic field. 
The scalar mode appears at the order $1/r$, the same as the scalar field,
while the electromagnetic mode appears at the finite order, the same as the
field $A_{A}$.

The zero modes are three arbitrary functions on the 2-sphere.  Thus, the kernel of $\Omega_{ss'}$ is three-dimensional in the subspace $r=\mathrm{const}.$, where its rank is zero.  The $r$-dependence   $v_{a}=\left( \frac{v(u,y)}{r},v^{A}(u,y)\right)$  will be inherited by the local parameters $\left( \xi ,\eta^{A}\right) =\left( \frac{\xi _{(1)}(u,y)}{r},\eta ^{A}(u,y)\right) $, as given by the matrix $\mathbf{Z}$ in \eqref{Zpar}.  

It is important to emphasize that the zero modes appear as multiplying
factors at each order in the expansion such that, when they are zero ($v=0$
and $v^{A}=0$), then all the modes are zero ($V=0$ and $V^{A}=0$). It means
that the only origin of the zero modes are the asymptotic ones, associated
with asymptotic symmetries.

Choosing $v_a=\left(\frac{v}{r},v_A\right)$, the above expressions identify the matrix $\mathbf{Z}$ in $V^{s}=Z^{sa}v_{a}$ up to $\mathcal{O}(\theta ^{3})$ terms as
\begin{equation}
\mathbf{Z}=\left(
\begin{array}{cc}
1-\frac{\theta ^{2}}{4e^{2}}\,\Delta  & \frac{1}{2e^{2}r}\left( -\epsilon \theta
\Delta _{B}+\frac{\theta ^{2}}{2}\,\hat{\Delta}_{B}\right)  \\
\frac{\epsilon \theta }{2}\,r\Delta ^{A}-\frac{\theta ^{2}}{4}\,r\left( \hat{\Delta}^{A}+\frac{\phi }{\sqrt{\gamma }}\,\epsilon ^{AB}\,\Delta _{B}\right)
& \delta _{B}^{A}-\frac{\epsilon \theta \phi }{2\sqrt{\gamma }}\,\epsilon
^{AC}\gamma _{CB}-\frac{\theta ^{2}}{4}\,\left( \frac{1}{e^{2}}\,\Delta _{\
B}^{A}+\frac{1}{2}\,\phi ^{2}\delta _{B}^{A}\right)
\end{array}
\right) \label{Z}
\end{equation}
The same matrix can be used to write out the first-class constraints $\Xi ^{a}=\chi _{s}Z^{sa}$, where now  $\Xi ^{a}$ are rows that act from the left side to the matrix $\mathbf{Z}$, arriving at 
\begin{equation}
\Xi =\left( 1-\frac{\theta ^{2}}{4e^{2}}\,\Delta \right) \hat{\chi}+\frac{\epsilon \theta }{2}\,r\Delta ^{A}\chi _{A}-\frac{\theta ^{2}}{4}\,r \left( \hat{\Delta}^{A}+\frac{\phi }{\sqrt{\gamma }}\,\epsilon ^{AB}\,\Delta _{B}\right)\chi _{A\,}+\mathcal{O}(\theta ^{3})\,,
\end{equation}
and
\begin{equation}
\Xi _{A}=\chi _{A}-\frac{\epsilon \theta }{2e^{2}r}\,\Delta _{A}\hat{\chi}+%
\frac{\epsilon \theta }{2}\sqrt{\gamma }\phi \,\epsilon _{AB}\chi ^{B}-\frac{%
\theta ^{2}}{8}\,\phi ^{2}\chi _{A}+\frac{\theta ^{2}}{4e^{2}}\,\left( \frac{1}{r%
}\,\hat{\Delta}_{A}\hat{\chi}-\chi ^{B}\Delta _{BA}\right) +\mathcal{O}(\theta ^{3})\,.
\end{equation}
It is important to notice that higher orders in $\theta$ are also higher orders in $1/r$. This can be deduced from the recursive formula  \eqref{recursive} where the coefficients of order $n-1$ are multiplied by high powers of the inverse radial coordinate, such that higher orders in the interaction parameters decrease faster to zero asymptotically. This feature of the zero modes is inherited by the first-class constraints, as well, because they are constructed from the same matrix $\mathbf{Z}$.
This fact will be important in the evaluation of the surface terms, such as charges, where all non-local higher-order terms will vanish on the asymptotic boundary.\medskip

To summarize, the first-class constraints are
\begin{equation}
\Xi \,,\quad \Xi _{A}\,,\quad \pi ^{u}\,,\quad \chi \,, \label{FirstCC}
\end{equation}
while the second-class constraints are the ones that remain of $\hat{\chi}$
and $\chi ^{A}$ when the first-class constraints are removed. Since they
interact non-linearly, it is difficult to decouple them explicitly.

\paragraph{Symmetry generators.} 

The smeared generators \eqref{smeared} for the first-class constraints are
\begin{eqnarray}
G[\alpha ] &=&\int \mathrm{d}^{3}x\,\left( \alpha \chi+\alpha _{u}\pi ^{u}\right) ,  \notag \\
K[\xi ] &=&\int \mathrm{d}^{3}x\,\xi\,\Xi \,, \label{gen}\\
S[\eta ] &=&\int \mathrm{d}^{3}x\,\eta _{A}\,\Xi^A  \,,\notag
\end{eqnarray}
where, again, the integrals include data at null infinity from  $u$ to $u\rightarrow -\infty $.
While $\alpha$ and $\alpha_u$ depend on spacetime coordinates $x^\mu$, to ensure that only the first-class constraints will enter in the above generators, we have to impose the following conditions on the local parameters $\eta_A$ and $\xi$,
\begin{equation}
    \xi=\frac{\xi_{(1)}(y)}{r}\,,\qquad \eta_A= \eta_A(y)\,. \label{eta,xi}
\end{equation}
As mentioned before, this $r$-dependence has been inherited from the zero modes of the symplectic matrix, while the $u$-independence is the requirement that quotients out trivial asymptotic transformations.

This is a generalization of the conditions \eqref{smear.S} and \eqref{smear.EM} imposed on the free scalar and the electromagnetic field, respectively. 

Let us prove that the constraints are indeed the first class, that is, that their PBs vanish weakly. As regards the form of the algebra, we will compute it later in terms of the conserved charges. 

For the generator $G[\alpha ]$, we find
\begin{eqnarray}
\left\{ G[\alpha _{1}],G[\alpha _{2}]\right\}  &=&0\,,  \notag \\
\left\{ G[\alpha ],K[\xi ]\right\}  &\approx &0\,, \\
\left\{ G[\alpha ],S[\eta ]\right\}  &\approx &0\,,  \notag
\end{eqnarray}
where the last two brackets follow from $\left\{ \chi ,\chi'_{s}\right\}
\approx 0$.  We remind the reader that `$\approx 0$' does not mean an approximate zero, but an exact zero on the constraint surface (weakly).  We look now at the brackets between the constraints $K[\xi ]$ and $S[\eta ]$. We get
\begin{eqnarray}
\left\{ K[\xi _{1}],K[\xi _{2}]\right\}  &\approx &\frac{\epsilon \theta ^{2}}{%
2e^{2}}\int \mathrm{d}^{3}x\sqrt{\gamma }r^{2}\xi _{1}\xi _{2}\left( -\Delta
^{A}\partial _{r}\Delta _{A}+\partial _{r}\Delta \right) +\mathcal{O}(\theta
^{3})\,,  \notag \\
\left\{ S[\eta _{1}],S[\eta _{2}]\right\}  &\approx &\frac{\epsilon \theta ^{2}}{2e^{4}}\,\int \mathrm{d}^{3}x\,\eta _{1A}\eta _{2B}\left[ \frac{1}{r}\,\left( \Delta ^{A}\epsilon ^{BC}-\,\epsilon ^{AC}\Delta ^{B}\right)
F_{rC}\right.   \notag \\
&&+\left. \sqrt{\gamma }\left( \partial _{r}\Delta ^{AB}-\Delta ^{A}\partial_{r}\Delta ^{B}\right) \right] +\mathcal{O}(\theta ^{3})\,, \\
\left\{ K[\xi ],S[\eta ]\right\}  &\approx &\int \mathrm{d}^{3}x\,\xi \eta
_{A}\left[ \frac{\theta }{e^{2}}\,\left( r\sqrt{\gamma }\,\partial _{r}\Delta^{A}-\epsilon ^{AB}\,F_{rB}\right) \right.   \notag \\
&&+\left. \frac{\epsilon \theta ^{2}}{2e^{2}}\,\sqrt{\gamma }\left( \phi \,F_{r}^{\ A}-r\partial _{r}\hat{\Delta}^{A}\right) \right]+\mathcal{O}(\theta ^{3})\,,  \notag
\end{eqnarray}
where we already accounted that $\partial _{r}\left( r\xi \right) =0$ and $
\partial _{r}\eta _{A}=0$, and also neglected all the boundary terms. Applying the identities \eqref{Delta} from Appendix \ref{Zmodes} leads to the vanishing of the brackets up to the $\mathcal{O}(\theta ^{3})$ order,
\begin{eqnarray}
\left\{ K[\xi _{1}],K[\xi _{2}]\right\}  &\approx &\mathcal{O}(\theta ^{3})\,,
\notag \\
\left\{ S[\eta _{1}],S[\eta _{2}]\right\}  &\approx &\mathcal{O}(\theta
^{3})\,, \\
\left\{ K[\xi ],S[\eta ]\right\}  &\approx &\mathcal{O}(\theta ^{3})\,,  \notag
\end{eqnarray}
as expected. The procedure explained in Section \ref{Formalism} guarantees that the
brackets will indeed vanish at all orders in $\theta$. We conclude that $G[\alpha]$, $K[\xi]$ and  $S[\eta]$ are indeed the generators corresponding to the first-class constraints \eqref{FirstCC}.

It is worth noticing that a more elegant way to write the smeared generators $K[\xi]$ and  $S[\eta]$ in \eqref{gen} in terms of the original primary constraints, 
\begin{eqnarray}
K[\xi ] =\int \mathrm{d}^{3}x\,\bar{\xi}\hat{\chi}\,, \qquad   S[\eta ] =\int \mathrm{d}^{3}x\,\bar{\eta}_{A}\,\chi ^{A}\,,
\end{eqnarray}%
where the local parameters are now determined from eq.~\eqref{Zpar}, namely,
they are transformed by the matrix $\mathbf{Z}$,
\begin{equation}
\bar{\xi}=\frac{\xi _{(1)}}{r}-\frac{\epsilon \theta }{2e^{2}r}\,\Delta
^{A}\eta _{A}-\frac{\theta ^{2}}{4e^{2}r}\,\left( \xi _{(1)}\,\Delta -\,\hat{%
\Delta}_{A}\eta ^{A}\right) +\mathcal{O}(\theta ^{3})\,,
\end{equation}%
and
\begin{eqnarray}
\bar{\eta}^{A} &=&\eta ^{A}+\frac{\epsilon \theta }{2}\,\left( \xi
_{(1)}\,\Delta ^{A}-\frac{\phi }{\sqrt{\gamma }}\,\epsilon ^{AB}\eta
_{B}\right) -\frac{\theta ^{2}\xi _{(1)}}{4}\,\left( \hat{\Delta}^{A}+\frac{%
\phi }{\sqrt{\gamma }}\,\epsilon ^{AB}\,\Delta _{B}\right)   \notag \\
&&-\frac{\theta ^{2}}{4}\,\left( \frac{1}{e^{2}}\,\Delta ^{AB}\eta _{B}+%
\frac{1}{2}\,\phi ^{2}\eta ^{A}\right) +\mathcal{O}(\theta ^{3})\,.
\end{eqnarray}
The non-trivial coupling between the scalar and electromagnetic field is transferred, through the matrix $\mathbf{Z}$ given by \eqref{Z}, from the constraints to the local parameters.  Nevertheless, the form of the generator remains unchanged because
\begin{eqnarray}
\left(
\begin{array}{cc}
\bar{\xi} & \bar{\eta}_{A}%
\end{array}%
\right) \left(
\begin{array}{c}
\hat{\chi} \\
\chi ^{A}%
\end{array}%
\right)  =\left(
\begin{array}{cc}
\xi  & \eta _{A}%
\end{array}%
\right) \mathbf{Z}\left(
\begin{array}{c}
\hat{\chi} \\
\chi ^{A}%
\end{array}%
\right)   =\left(
\begin{array}{cc}
\xi  & \eta _{A}%
\end{array}%
\right) \left(
\begin{array}{c}
\Xi  \\
\Xi ^{A}%
\end{array}%
\right) \,,
\end{eqnarray}
where $\xi$ and $\eta$ are given by \eqref{eta,xi}.

\paragraph{Symmetry transformations.} 

The above functionals generate the transformations $\delta _{\alpha }=\{\quad,G[\alpha ]\,\}$, $\delta
_{\eta }=\{\quad ,S[\eta ]\,\}$ and  $\delta _{\xi }=\{\quad ,K[\xi ]\,\}$ acting on any function on the phase space. Under $G$, the fields transform according to the usual $\mathrm{U}(1)$ transformation law,
\begin{equation}
\delta _{\alpha }A_{\mu }  =-\delta _{\mu }^{i}\,\partial_{i}\alpha +\delta^u_{\mu}\,\alpha _{u}\,. \label{tr.0}
\end{equation}

The asymptotic symmetries generated by $K$ and $S$ are non-local and couple the scalar and electromagnetic fields.  Using the non-vanishing auxiliary PBs with $\chi_s$, 
\begin{equation}
\begin{array}{llll}
\left\{ \phi ,\hat{\chi}'\right\}  & =\epsilon \delta
^{(3)}\,,\medskip  & \left\{ A_A,\chi ^{\prime B}\right\}  & =\epsilon
\delta _A^B\delta ^{(3)}\,, \\
\left\{ \pi ,\hat{\chi}'\right\}  & =-\sqrt{\gamma }\,r^{\prime
2}\partial _{r}\delta ^{(3)}\,,\medskip  & \left\{ \pi ^{r},\chi ^{\prime
A}\right\}  & =\sqrt{\gamma'}\,\sigma ^{\prime AB}\partial
_{B}\delta ^{(3)}\,, \\
\left\{ \pi ,\chi ^{\prime A}\right\}  & =-\dfrac{\epsilon \theta }{e^{2}}%
\,\epsilon ^{AB}F_{rB}\,\delta ^{(3)}\,,\qquad & \left\{ \pi ^{A},\chi ^{\prime B}\right\}  & =-\sqrt{\gamma }\,\sigma ^{\prime BA}\partial _{r}\delta
^{(3)}\,,
\end{array}
\end{equation}
we find that non-trivial transformations for the fields are
\begin{eqnarray}
\delta _{\xi }\phi  &\approx &\epsilon \xi \left( 1-\frac{\theta ^{2}}{4e^{2}}%
\,\Delta \right) +\mathcal{O}(\theta ^{3})\,,  \notag \\
\delta _{\xi }A_{A} &\approx &\frac{\theta }{2}\,r\xi \,\left[ \Delta _{A}-%
\frac{\epsilon \theta }{2}\,\left( \hat{\Delta}_{A}+\sqrt{\gamma }\,\phi\,\epsilon
_{AB} \Delta ^{B}\right) \right] +\mathcal{O}(\theta ^{3})\,, \label{tr.1}
\end{eqnarray}
and also
\begin{eqnarray}
\delta _{\eta }\phi  &\approx &\frac{\theta }{2e^{2}r}\,\eta ^{A}\left(
-\Delta _{A}+\frac{\epsilon \theta }{2}\,\hat{\Delta}_{A}\right) +\mathcal{O}%
(\theta ^{3})\,,  \notag \\
\delta _{\eta }A_{A} &\approx &\epsilon \eta _{A}\left( 1-\frac{\theta ^{2}}{%
8}\,\phi ^{2}\right) -\frac{\theta }{2}\,\eta ^{B}\left( \sqrt{\gamma }\phi
\,\epsilon _{AB}+\frac{\epsilon \theta }{2e^{2}}\,\Delta _{AB}\right) +%
\mathcal{O}(\theta ^{3})\,.  \label{tr.2}
\end{eqnarray}
As discussed before, all $\mathcal{O}(\theta ^{3})$ terms tend very fast to zero as the radius approaches the boundary, so the $\theta$-expansion becomes $r^{-1}$-expansion for large values of the radii. 
As regards the momenta, we get the following transformations,
\begin{eqnarray}
\delta _{\xi }\pi  &\approx &-\sqrt{\gamma }r\xi \,\left( 1-\frac{\theta ^{2}}{4e^{2}}\,\Delta \right)
-\frac{\theta ^{2}}{4e^{2}}\,r\xi \epsilon ^{AB}\Delta _{A}F_{rB}+\mathcal{O}%
(\theta ^{3})\,,  \notag \\
\delta _{\xi }\pi ^{r} &\approx &\partial _{A}N^{A}+\mathcal{O}(\theta
^{3})\,, \\
\delta _{\xi }\pi ^{A} &\approx &-\partial _{r}N^{A}+\mathcal{O}(\theta
^{3})\,,  \notag
\end{eqnarray}%
as well as
\begin{eqnarray}
\delta _{\eta }\pi  &\approx &-\sqrt{\gamma }\partial _{r}\left[ \frac{\theta }{2e^{2}}\,r\eta ^{A}\left( -\epsilon \Delta _{A}+\frac{\theta }{2}\,
\hat{\Delta}_{A}\right) \right]  \notag \\
&&+\sqrt{\gamma }\,\frac{\theta }{e^{2}}\left(
-\epsilon \,\epsilon ^{AB}+\frac{\theta }{2}\,\gamma ^{AB}\phi \right) \eta
_{A}F_{rB}+\mathcal{O}(\theta ^{3})\,,  \notag \\
\delta _{\eta }\pi ^{r} &\approx &\partial _{A}M^{A}+\mathcal{O}(\theta
^{3})\,,  \label{tr.3} \\
\delta _{\eta }\pi ^{A} &\approx &-\partial _{r}M^{A}+\mathcal{O}(\theta
^{3})\,,  \notag
\end{eqnarray}
where we denoted
\begin{eqnarray}
N^{A} &\equiv&\frac{\theta }{2}\,\sqrt{\gamma }\, r\xi \,\sigma ^{BA}\left(
\epsilon \Delta _{B}-\frac{\theta }{2}\,\hat{\Delta}_{B}-\frac{\theta }{2}\,
\sqrt{\gamma }\,\phi \epsilon _{BC}\,\Delta ^{C}\right)  \,,  \notag \\
M^{A} &\equiv&\sqrt{\gamma }\,\eta ^{C}\left[ \left( 1-\frac{\theta ^{2}}{8}
\,\phi ^{2}\right) \gamma _{BC}+\frac{\theta }{2}\left( \epsilon \sqrt{\gamma }\phi \,\epsilon _{FB}-\frac{\theta }{2e^{2}}\,\Delta _{BC}\right) 
\right] \sigma ^{BA}\,.  \label{M,N}
\end{eqnarray}

We note that, when $\theta =0$, the new transformations that appear only on the null infinity given by \eqref{tr.1}--\eqref{tr.2} are just the shift transformations of the leading components of the fields. Those transformations are asymptotic because they are not symmetries in the bulk of the spacetime. Interestingly, the shift symmetry $\delta\phi=\mathrm{const.}$, which is the true bulk symmetry, has not been detected by this procedure, because the Hamiltonian method sees only symmetries whose parameters depend on some coordinates. 
The addition of the terms with $\theta \neq 0$ couples non-trivially $\phi$ and the angular components of the electromagnetic field, $A_A$.

\paragraph{Conserved charges.}

Conserved charges in Hamiltonian formalism are boundary terms in the improved smeared generators. According to the Regge-Teitelboim method \cite{Regge:1974zd}, we define the
improved generators as
\begin{eqnarray}
G_{K}[\xi ] &=&K[\xi ]+Q_{K}[\xi ]\,,\notag\\
G_{Q}[\alpha ] &=&G[\alpha ]+Q[\alpha ]\,,  \\
G_{S}[\eta ] &=&S[\eta ]+Q_{S}[\eta ]\,,   \notag
\end{eqnarray}%
where $Q$, $Q_{S}$ and $Q_{K}$ are the boundary terms that make the generators differentiable. To compute them, we vary the generators and identify the following boundary terms,
\begin{equation}
\delta G[\alpha ]=\text{(bulk term)}+\oint \mathrm{d}^{2}y\,\alpha \delta
\pi ^{r}\,,
\end{equation}%
and also
\begin{eqnarray}
\delta K[\xi ] &\approx &\text{(bulk term)}+\oint \mathrm{d}^{2}y\,\xi \,\left[ -r^{2}\left( 1-\frac{\theta ^{2}}{4e^{2}}\,\Delta \right) \sqrt{\gamma } \,\delta \phi -\frac{\epsilon \theta }{2}\,r\Delta _{A}\sqrt{\gamma }\,\sigma
^{AB}\delta A_{B}\right.   \notag \\
&&\qquad \qquad \qquad +\left. \frac{\theta ^{2}}{4}\,r\left( \hat{\Delta}_{A}+\phi \,\sqrt{\gamma }\epsilon _{AC}\,\Delta ^{C}\right) \sqrt{\gamma } \,\sigma ^{AB}\delta A_{B}\right] +\mathcal{O}(\theta ^{3})\,,   \\
\delta S[\eta ] &\approx &\text{(bulk term)}+\oint \mathrm{d}^{2}y\,\eta ^{A}
\left[ -\sqrt{\gamma }\,\sigma _{AB}\delta A^{B}+\frac{\epsilon \theta }{2e^{2}}\,r\Delta _{A}\sqrt{\gamma }\,\delta \phi -\frac{\epsilon \theta }{2}\sqrt{\gamma }\phi \,\epsilon _{AB}\sqrt{\gamma }\,\sigma ^{BC}\delta A_{C}\right. \notag
\\
&&\qquad \qquad \qquad +\left. \frac{\theta ^{2}}{8}\,\phi ^{2}\sqrt{\gamma } 
\,\sigma _{AB}\delta A^{B}+\frac{\theta ^{2}}{4e^{2}}\,\left( -r\hat{\Delta}_{A}\sqrt{\gamma }\,\delta \phi +\sqrt{\gamma }\,\sigma ^{BC}\delta A_{C}\Delta _{BA}\right) \right] +\mathcal{O}(\theta ^{3})\,.  \notag
\end{eqnarray}
In the above computation, we used $\int \mathrm{d}^{3}x\,\partial _{B}\left(
\sqrt{\gamma }V^{B}\right) =0$, assuming that the 2-sphere is smooth. Because the boundary terms are asymptotic, taking the limit $r\rightarrow
\infty $ allows to neglect the non-local terms $\Delta _{A}=\mathcal{O}(r^{-2})$, $\hat{\Delta}_{A}=\mathcal{O}(r^{-3})$, $\Delta =\mathcal{O}(r^{-4})$  and $\Delta _{AB}=\mathcal{O}(r^{-4})$. Also some scalar fields
vanish due to $\phi =\mathcal{O}(r^{-1})$, leaving the only finite contributions
\begin{eqnarray}
-r^{2}\xi \,\sqrt{\gamma }\,\delta \phi  &=&-\xi _{(1)}\,\sqrt{\gamma }
\,\delta \phi _{(1)}+\mathcal{O}(r^{-2})\,,  \notag \\
-\sqrt{\gamma }\,\sigma _{AB}\eta ^{A}\delta A^{B} &=&-\frac{1}{e^{2}}\,%
\sqrt{\gamma }\eta ^{A}\delta A_{A}+\mathcal{O}(r^{-1})\,,
\end{eqnarray}%
to get
\begin{eqnarray}
\delta Q_{K}[\xi ] &=&r^{2}\oint \mathrm{d}^{2}y\,\sqrt{\gamma }\xi \delta
\phi +\mathcal{O}(\theta ^{3})\,, \notag  \\
\delta Q[\alpha ] &=&-\oint \mathrm{d}^{2}y\,\alpha \delta \pi ^{r}\,,
\\
\delta Q_{S}[\eta ] &=&\frac{1}{e^{2}}\,\oint \mathrm{d}^{2}y\,\sqrt{\gamma }%
\eta ^{A}\delta A_{A}+\mathcal{O}(\theta ^{3})\,.  \notag
\end{eqnarray}
Besides, as discussed previously in the paragraph before eq.~\eqref{FirstCC}, the terms of higher order in $\theta$ are also higher order in $1/r$, so 
we can take that $\mathcal{O}(\theta ^{3})$ vanish asymptotically.

Assuming that the parameters are field-independent, the charges can be integrated as
\begin{eqnarray}
Q_{K}[\xi ] &=&r^2\oint \mathrm{d}^{2}y\,\sqrt{\gamma }\,\xi \phi \,,\notag
 \\
Q[\alpha ] &=&-\oint \mathrm{d}^{2}y\,\alpha \pi ^{r}\,, \label{EMP charges}  \\
Q_{S}[\eta ] &=&\frac{1}{e^{2}}\,\oint \mathrm{d}^{2}y\,\sqrt{\gamma }\eta
^{A} A_{A}\,.  \notag
\end{eqnarray}
All the expressions are finite when evaluated on the 2-sphere.

Obtained charges do not depend on the coupling constant $\theta $, and they are the same charges as the ones corresponding to the free scalar and electromagnetic fields. More precisely, the scalar charge $Q_K$ coincides with the one of the Section \ref{Scalar} given by eq.~\eqref{charge.s}, and the gauge charge $Q$ also coincides with the gauge charge in \eqref{charges.EM} from Section \ref{EM+P}. 

However, the coupling with the Pontryagin term and the scalar field considered in this section does not reproduce the same electromagnetic charges computed in the previous Section \ref{EM+P} with only the Pontryagin contribution, see expression for $Q_S$ in \eqref{charges.EM}. 
A difference is in the coupling. While in the previous section the interaction was introduced through the constant $\theta$, the interaction here is defined through  $\theta\phi = \mathcal{O}(r^{-1})$, and it has the faster fall-off which results in decoupling the charges on the boundary. This is why obtained $Q_S$ corresponds to the free Maxwell field of ref.~\cite{Gonzalez:2023yrz}.

\paragraph{Charge algebra.}

Obtained transformations are also symmetries of the symplectic structure. Namely, the phase space is equipped with the canonical symplectic form
\begin{equation}
\omega =\int \mathrm{d}^{3}x\,\left( \delta \pi \wedge \delta \phi +\delta
\pi ^{\mu }\wedge \delta A_{\mu }\right) \,,
\end{equation}%
and symmetries discovered previously can be generated by vector fields producing transformations \eqref{tr.0}--\eqref{tr.3} through the contractions $i_{X_{\alpha }}\delta =\delta _{\alpha }$, $i_{X_{\xi }}\delta
=\delta _{\xi }$ and $i_{X_{\eta }}\delta =\delta _{\eta }$. They are isometries of the symplectic manifold because of the identity $i_{X}\omega
=-\delta Q$ satisfied by the charges \eqref{EMP charges}, which implies $\mathscr{L}_{X}\omega=0$.

The algebra of these charges is computed as a negative double contraction of the symplectic form,
\begin{equation}
\left\{ Q[\alpha _{1}],Q[\alpha _{2}]\right\}^{\ast } =-\int \mathrm{d}^{3}x\,\left( \delta _{\alpha _{2}}\pi \delta _{\alpha _{1}}\phi -\delta _{\alpha _{1}}\pi
\delta _{\alpha _{2}}\phi +\delta _{\alpha _{2}}\pi ^{\mu }\delta _{\alpha_{1}}A_{\mu }-\delta _{\alpha _{1}}\pi ^{\mu }\delta _{\alpha _{2}}A_{\mu
}\right) \,,
\end{equation}
and similarly for all other brackets. Nonetheless, because the charges \eqref{EMP charges} are the ones of the free scalar and electromagnetic fields without the Pontryagin term, we expect that they satisfy the same algebra as the one found for the free fields. This is consistent with the fact that all coupling terms in the algebra vanish at the asymptotic infinity. Namely, all non-local higher-order $\theta $-terms decrease to zero very fast, while the linear term vanishes due to the fall-off of the scalar.

To illustrate this, we will show explicitly that the bracket between the scalar charge and the electromagnetic $\mathrm{U}(1)$ generators indeed vanishes. This is seen from the double contraction of the symplectic form by the corresponding vector fields,
\begin{eqnarray}
i_{X_{\xi }}i_{X_{\alpha }}\omega 
=-\oint \mathrm{d}^{2}y\,N^{A}\partial _{A}\alpha +\mathcal{O}(\theta
^{3})=\mathcal{O}(\theta ^{3})\,,
\end{eqnarray}
where the vanishing result is because $\lim_{r\rightarrow \infty }N^{A}=
\mathcal{O}(\theta ^{3})$, as it can be seen from \eqref{M,N} and
\eqref{Delta}. Furthermore, we already saw that the $\mathcal{O}(\theta ^{3})$ terms vanish on the asymptotic 2-sphere because of the fast fall-off of the non-local terms. Therefore, we can take $\mathcal{O}(\theta
^{3})\rightarrow 0$ when $r\rightarrow \infty $, and indeed obtain the exact
result $\left\{ Q[\alpha ],Q_{K}[\xi ]\right\} =0$. Similar computation can
be done for all other brackets, and they all vanish except one. Namely,
the above procedure leads to
\begin{equation}
i_{X_{\eta }}i_{X_{\alpha }}\omega =-\oint \diff^{2}y\,M^{A}\partial
_{A}\alpha =-\frac{1}{e^{2}}\oint \diff^{2}y\,\sqrt{\gamma }\,\eta
^{A}\partial _{A}\alpha \,,
\end{equation}%
where $M^{A}$ given by \eqref{M,N} has one finite term at the
infinity, $\lim_{r\rightarrow \infty }M^{A}=\frac{1}{e^{2}}\sqrt{\gamma }\,\eta ^{A}$. This term will give rise to the central charge in the algebra.

In sum, we find that the algebra of charges is Abelian, with only one PB not vanishing, featuring the central extension
\begin{equation}
\left\{ Q[\alpha ],Q_{S}[\eta ]\right\}^{\ast } =C[\alpha ,\eta ]=\frac{1}{e^{2}}
\oint \mathrm{d}^{2}y\,\sqrt{\gamma }\,\eta^{A}\partial
_A\alpha \,.
\end{equation}
A difference between this charge and the one found in the previous section, see eq.~\eqref{C.EM}, is that the coupling term here vanishes because of the asymptotic behavior of the scalar field, $\theta \phi \rightarrow 0$. Therefore, the above charge is
the one found in the Maxwell's electrodynamics \cite{Gonzalez:2023yrz}.\medskip

We found that the coupling between the scalar and electromagnetic fields
interacting through the Pontryagin term makes the bulk dynamics of the
theory richer, but the asymptotic one becomes simpler than the one described
by the scalar and electromagnetic theories with the Pontyagin topological
invariant.

In particular, the theory with the scalar coupling does not exhibit the
electric-magnetic duality.

\section{Discussion}
\label{Discussion}

We have examined the asymptotic symmetries of massless theories using the Dirac formalism in the null foliation of Minkowski space-time, providing insights into their structure at null infinity. In general, large gauge
charges can also be analyzed in the context of massive theories at timelike infinity, as discussed in QED in \cite{Herdegen:1995nf} and more recently in \cite{Campiglia:2015qka}. It turns out that they also have null infinity contribution \cite{Campiglia:2017mua}. A more general analysis of the role of asymptotically non-trivial gauge transformations, for both massive and massless particles, was discussed in \cite{Guendelman:1989if}. However, the focus of our work is
purely on massless theories.

As pointed out previously in \cite{Gonzalez:2023yrz}, massless theories in the null foliation have additional gauge generators not present in time-like foliation, responsible for new soft charges. To better understand the nature of these constraints, we extended the Dirac procedure for constrained systems by introducing  the matrix $\mathbf{Z}$, which encodes information
about the zero modes \eqref{zv.modes} of the symplectic matrix $\Omega $ constructed from the PBs of constraints in the theory. This matrix plays a crucial role in our discussion, as it is used to construct arbitrary Hamiltonian multipliers \eqref{Zlambda} that introduce ambiguity in evolution, identify the first-class constraints in the theory \eqref{FCC}, obtain associated symmetry generators \eqref{smeared} and soft charges, and
find the form of the local parameters of the symmetry \eqref{Zpar}.

In particular, we applied this formalism to the scalar-Maxwell theory described by the action \eqref{s.EM}.

We started by exploring the sector of the theory with $A_{\mu }=0$, corresponding to the scalar field in the Bondi background. We found that the scalar field possesses nontrivial asymptotic symmetries, consisting of an
infinite number of Abelian symmetries with the parameters $\xi _{(1)\ell m}$ when expanded in spherical harmonics. Its soft charges coincide with those obtained in \cite{Campiglia:2017dpg} via the soft theorem. This intriguing result, that the boundary charges are not related to the gauge symmetries of the bulk, has also been observed in \cite{Freidel:2018fsk}. Nonetheless, the soft scalar charges can be seen as large
gauge charges in a dual 2-form field formulation \cite{Campiglia:2018see,Francia:2018jtb}. 

We then analyzed the sector of the theory with $\phi =\theta =\mathrm{const.}$,  describing Maxwell's electromagnetic theory supplemented by the topological Pontryagin term. Pure electromagnetism was first discussed using the same formalism in \cite{Gonzalez:2023yrz}, finding asymptotic shift symmetries of the fields $A_{A}(y)$ at the null boundary. They can be interpreted as large gauge transformations of the dual magnetic potential
\cite{Strominger:2015bla}, and are related to the magnetic charge $Q_{\mathrm{mag}}$ in the theory \cite{Campiglia:2016hvg}. They generate the
magnetic soft photon theorem \cite{Strominger:2015bla} in a similar way as the gauge transformations are related to the electric charge, $Q_{\mathrm{el}}$, and generate the usual Weinberg's soft photon theorem. 
In anti-de Sitter space, asymptotic symmetries and their associated charges in electromagnetism have been discussed in \cite{Esmaeili:2019mbw}.
Coupled with gravity, non-stationary Maxwell fields contribute to the charges for asymptotic Lorentz symmetries at the null infinity \cite{Bonga:2019bim}. In the context of the convolutional double copy, the asymptotic symmetries of Maxwell fields lead to double-copy supertranslations, and the necessity of logarithmic terms in the radial expansion to obtain non-vanishing asymptotic charges \cite{Ferrero:2024eva}.

When the Pontryagin term is added, these symmetries remain, but the theory exhibits the electric-magnetic duality $\mathrm{SL}(2,\mathbb{R})$ between the boundary charges. This is another
example of a system where the symmetry is not realized as a bulk duality but it is purely asymptotic \cite{Freidel:2018fsk,Nande:2017dba}. However, adding a second scalar field in the theory \eqref{s.EM}, it is possible to extend the symmetry to the spacetime $\mathrm{SL}(2,\mathbb{R})$-duality \cite{Fuentealba:2024mor}.
In \cite{Nande:2017dba}, the quantized electric and magnetic charges live on the charge lattice, such that an electromagnetic duality symmetry becomes $\mathrm{SL}(2,\mathbb{Z})$.
The electric-magnetic duality exists only with the Pontryagin term, and it vanishes when $\theta =0 $.

The charge algebra \eqref{alg.EM} possesses a non-trivial central extension  \eqref{C.EM} that translates to non-trivial PBs between $Q_{\mathrm{mag}}$ and $Q_{\mathrm{el}}$ equal to the electric-magnetic central charge $C_{\mathrm{EM}}$ \eqref{C_EM}, which contains an additional term compared to the one in \cite{Freidel:2018fsk}. The existence of $C_{\mathrm{EM}}$ could explain why a magnetic monopole cannot be observed. Central charge in electromagnetic duality can also be obtained in the context of $p$-form theories \cite{Geiller:2021gdk}.

A remaining problem yet to be understood is how the algebra \eqref{algebraTheta} depends on the parameter $\theta $. Although the total charge is invariant under $\mathrm{SL}(2,\mathbb{R})$ symmetry, it is not clear whether the
corresponding Abelian charge algebra shares this property, that is, whether different choices of parameters lead to equivalent central extensions.

Another, maybe related, question to clarify is the influence of the function
$k(y)$ that appears in \eqref{integrated}  when $\theta =0$ or in \eqref{integrated-theta} when $\theta \neq 0$. It can be computed using eq.~\eqref{dotPi} under the integral $\int \mathrm{d}^{2}y\mathrm{d}u  \, \partial _{u}$. For example, when $\epsilon =1$, the Hamiltonian $\mathrm{U}(1)$ charge can be written in terms of $A_{A}$ in a gauge-invariant way as $$Q[\alpha ]=-\frac{1}{e^{2}}\int \mathrm{d}^{2}y\,\sqrt{\gamma }\,\alpha
\nabla ^{A}\left[ A_{A}(u\rightarrow +\infty ,y)-A_{A}(u\rightarrow -\infty
,y)\right]\,,$$ such that $k(y)=-\frac{1}{e^{2}}\,\sqrt{\gamma }\,\alpha
\nabla ^{A}A_{(0)A}(u\rightarrow +\infty ,y)\,,$ and similarly for $\epsilon =-1$.  Due to its field dependence, the quantity $k(y)$ appears as a new
canonical variable in the charges. Furthermore, dependence of the electric
charge from $\Delta A_{A}=A_{A}(u\rightarrow +\infty ,y)-A_{A}(u\rightarrow
-\infty ,y)$ is a manifestation of the electromagnetic memory effect \cite{He:2014cra}, and it would be interesting to discuss how it would affect central extensions using present formalism.

In the full interacting theory where both $A_{\mu }$ and $\phi $ are dynamical, electromagnetic duality is broken, but asymptotic symmetry remains. This theory has a more complicated dynamic structure in the bulk due to scalar-vector interaction, but it is asymptotically simpler because the scalar and electromagnetic fields become decoupled.

Finally, it is worth mentioning that the interacting theory has been analyzed only when $\theta $ is arbitrary, excluding the (anti-)self-dual point $\theta \neq \pm 1$, because it could change the boundary behavior of
the electromagnetic field. For the (anti-)self-duality $^* F_{\mu \nu }=\pm F_{\mu \nu }$ to be realized, the 2-sphere has to be projected to the complex plane, which has a different metric signature. This is one of the open questions that we plan to address soon.

\section*{Acknowledgments}

We would like to thank Miguel Campiglia, Milutin Blagojevi\'c, Laurent Freidel,  Marc Henneaux, Kevin Nguyen, \DJ or\dj e Mini\'c, Alfredo Pérez, Marios Petropoulos, and Daniele Pranzetti for useful discussions. We also thank the anonymous referee for their valuable feedback and insightful suggestions, which have helped improve this manuscript. This work has been funded in part by Anillo Grant ANID/ACT210100 \textit{Holography and its Applications to High Energy Physics, Quantum Gravity and Condensed Matter Systems} and FONDECYT Regular Grants 1221920, 1230492, 1230853, and 1231779.

\appendix

\section{Zero modes of the symplectic matrix} \label{Zmodes}

Here we solve the equation of zero modes of the matrix $\Omega_{ss'}$ constructed in Section \ref{EM with scalar}, given by the system of partial linear differential equations \eqref{PDE} in $V^{s}$, using the method of successive approximations. 

Expanding $V^s$ in different orders $V^s_n$ ($n=0,1 \cdots$) of small $\theta $, the system reduces to the recursive equations \eqref{recursive}.  In particular, at the zeroth order in $\theta $, these equations solve the leading order of the vector $V^{s}$,
\begin{eqnarray}
0 &=&\partial _{r}(rV_{0})\quad \Rightarrow \quad V_{0}=\frac{v_{0}(u,y)}{r}%
\,,  \notag \\
0 &=&\partial _{r}V_{0}^{A}\quad \quad \Rightarrow \quad
V_{0}^{A}=v_{0}^{A}(u,y)\,.
\end{eqnarray}%
At the linear order in $\theta $, we obtain
\begin{eqnarray}
0 &=&\partial _{r}(rV_{1})+\frac{\epsilon }{2e^{2}r\sqrt{\gamma }}\,\epsilon
^{AB}F_{rB}\,V_{0A}\,,  \notag \\
0 &=&\partial _{r}V_{1}^{A}-\frac{\epsilon }{2\sqrt{\gamma }}\,\epsilon
^{AB}F_{rB}\,V_{0}+\frac{\epsilon }{2\sqrt{\gamma }}\,\epsilon ^{AB}\partial
_{r}\phi \,V_{0B}\,,
\end{eqnarray}
and the next-to-leading order of the vector is solved as
\begin{eqnarray}
V_{1} &=&\frac{v_{1}(u,y)}{r}-\frac{\epsilon }{2e^{2}r}\,\Delta ^{A}v_{0A}\,,
\notag \\
V_{1}^{A} &=&v_{1}^{A}(u,y)+\frac{\epsilon }{2}\,v_{0}\,\Delta ^{A}-\frac{%
\epsilon \phi }{2\sqrt{\gamma }}\,\epsilon ^{AB}v_{0B}\,.
\end{eqnarray}
To simplify the expressions, we introduced the non-local functions $\Delta^{A}$ given by eqs.~\eqref{nonlocal}. Finally, the differential equations at the quadratic order in $\theta $ take the form
\begin{eqnarray}
0 &=&\partial _{r}(rV_{2})+\frac{\epsilon }{2e^{2}r\sqrt{\gamma }}\,\epsilon
^{AB}F_{rB}\,V_{1A}\,,  \notag \\
0 &=&\partial _{r}V_{2}^{A}-\frac{\epsilon }{2\sqrt{\gamma }}\,\epsilon
^{AB}F_{rB}\,V_{1}+\frac{\epsilon }{2\sqrt{\gamma }}\,\epsilon ^{AB}\partial
_{r}\phi \,V_{1B}\,,
\end{eqnarray}%
leading to the solution
\begin{eqnarray}
V_{2} &=&\frac{v_{2}(u,y)}{r}-\frac{1}{2e^{2}r}\,\left( \epsilon \Delta
^{A}v_{1A}+\frac{1}{2}\,v_{0}\,\Delta -\frac{1}{2}\,\hat{\Delta}%
_{A}v_{0}^{A}\right) \,,  \notag \\
V_{2}^{A} &=&v_{2}^{A}(u,y)+\frac{\epsilon }{2}\,v_{1}\Delta ^{A}-\frac{1}{%
4e^{2}}\,\Delta ^{AB}v_{0B}-\frac{v_{0}}{4\sqrt{\gamma }}\,\phi \epsilon
^{AB}\Delta _{B} \\
&&-\,\frac{v_{0}}{4}\,\hat{\Delta}^{A}-\frac{\epsilon \phi }{2\sqrt{\gamma }}%
\,\epsilon ^{AB}v_{1B}-\frac{1}{8}\,\phi ^{2}v_{0}^{A}\,,  \notag
\end{eqnarray}%
where again the non-local functions $\Delta $, $\hat{\Delta}_{A}$ and $\Delta ^{AB}$ are given by \eqref{nonlocal}.
Collecting the obtained terms together, we find  the approximate form of the zero modes as
\begin{equation}
V=\frac{v}{r}-\frac{\epsilon \theta }{2e^{2}r}\,\Delta ^{A}v_{A}-\frac{%
\theta ^{2}}{4e^{2}r}\,\left( v\,\Delta -\,\hat{\Delta}_{A}v^{A}\right) +
\mathcal{O}(\theta ^{3})\,,
\end{equation}%
and
\begin{eqnarray}
V^{A} &=&v^{A}+\frac{\epsilon \theta }{2}\,\left( v\,\Delta ^{A}-\frac{\phi
}{\sqrt{\gamma }}\,\epsilon ^{AB}v_{B}\right) -\frac{\theta ^{2}v}{4}%
\,\left( \hat{\Delta}^{A}+\frac{\phi }{\sqrt{\gamma }}\,\epsilon
^{AB}\,\Delta _{B}\right)   \notag \\
&&-\frac{\theta ^{2}}{4}\,\left( \frac{1}{e^{2}}\,\Delta ^{AB}v_{B}+\frac{1}{%
2}\,\phi ^{2}v^{A}\right) +\mathcal{O}(\theta ^{3})\,,
\end{eqnarray}%
where we denoted $v^{s}=v_{0}^{s}+\theta v_{1}^{s}+\theta ^{2}v_{2}^{s}+
\mathcal{O}(\theta ^{3})$. Choosing $v_{a}=\left( \frac{v}{r},v_{A}\right) $%
, the matrix $\mathbf{Z}=[Z^{sa}]$ can be read off directly from $V^{s}=Z^{sa}v_{a}$
up to $\mathcal{O}(\theta ^{3})$ terms, as given by \eqref{Z}.\medskip

In the above computation, we used the following properties of the non-local functions \eqref{nonlocal}:
\begin{equation}
\begin{array}{llll}
\partial _{r}\Delta ^{A} & =\dfrac{1}{r\sqrt{\gamma }}\,\epsilon^{AB}F_{rB}\,,\medskip  & \partial _{r}\hat{\Delta}_{A} & =\dfrac{1}{r}\,\phi F_{rA}\,, \\
\partial _{r}\Delta  & =\dfrac{1}{r\sqrt{\gamma }}\,\epsilon ^{AB}\,\Delta
_{A}F_{rB}\,,  \qquad& \partial _{r}\Delta ^{AB} & =\dfrac{1}{r\sqrt{\gamma }}
\,\epsilon ^{AC}\,F_{rC}\,\Delta ^{B}\,.
\end{array}
\label{Delta}
\end{equation}

\bibliography{bibAsym}
\bibliographystyle{utphys}

\end{document}